\documentclass[natbib]{svjour3}             
\smartqed  
\usepackage{graphicx}
\usepackage{aps-bibstyle}  
\begin{document}

\title{Oscillations and waves in solar spicules
}



\author{T.V. Zaqarashvili        \and
        R. Erd\'elyi 
}

\authorrunning{Zaqarashvili and Erd\'elyi} 

\institute{T.V. Zaqarashvili \at Abastumani Astrophysical Observatory at the
              Faculty of Physics and Mathematics, I. Chavchavadze State
              University, Chavchavadze Ave. 32, Tbilisi 0179, Georgia \\
              Tel.: +995-32-294714\\
              Fax: +995-32-220009\\
              \email{temury.zaqarashvili@iliauni.edu.ge}           
           \and
           R. Erd\'elyi \at
              Solar Physics \& Space Plasma Research Centre (SP$^2$RC) \\
              Department of Applied Mathematics\\
              University of Sheffield \\
              Sheffield S3 7RH, UK \\
              \email{robertus@sheffield.ac.uk}
}

\date{Received: date / Accepted: date}

\maketitle

\begin{abstract}
Since their discovery, spicules have attracted increased attention
as energy/mass bridges between the dense and dynamic photosphere and
the tenuous hot solar corona. Mechanical energy of photospheric
random and coherent motions can be guided by magnetic field lines,
spanning from the interior to the upper parts of the solar
atmosphere, in the form of waves and oscillations. Since spicules
are one of the most pronounced features of the chromosphere, the
energy transport they participate in can be traced by the
observations of their oscillatory motions. Oscillations in spicules
have been observed for a long time. However the recent
high-resolutions and high-cadence space and ground based facilities
with superb spatial, temporal and spectral capacities brought new
aspects in the research of spicule dynamics. Here we review
the progress made in imaging and spectroscopic observations of waves
and oscillations in spicules. The observations are accompanied by
a discussion on theoretical modelling and interpretations of these
oscillations. Finally, we embark on the recent developments made
on the presence and role of Alfv\'en and kink waves in spicules.
We also address the extensive debate made on the Alfv\'en versus
kink waves in the context of the explanation of the observed
transverse oscillations of spicule axes.

\end{abstract}

\section{Introduction}
\label{sec:intro}

The rapid rise of plasma temperature up to 1 MK from the solar
photosphere towards the corona is still an unresolved problem in solar
physics. It is clear that the mechanical energy of sub-photospheric
motions is transported somehow into the corona, where it may be
dissipated leading to the heating of the ambient plasma. A possible
scenario of energy transport is that the convective motions and
solar global oscillations may excite magnetohydrodynamic (MHD) waves in
the photosphere, which may then propagate through the chromosphere carrying
relevant energy into the corona. It is of great desire that the energy
transport process(es) can be tracked by observational evidence of
the oscillatory phenomena in the chromosphere. For a detailed discussion
about MHD wave heating and heating diagnostics in the solar atmosphere
see the recent work by \cite{tar09}.

Much of the radiation from the upper chromosphere originates in {\it
spicules}, which are grass-like spiky features seen in chromospheric
spectral lines at the solar limb (see Fig 1). These abundant and
spiky features in the chromosphere were discovered by \cite{sec77}
and were named "spicules" by \cite{rob45}. \cite{bec68,bec72,ste00}
dedicated excellent reviews to summarizing the observational and
theoretical views about spicules at that time. Since these reviews,
many observational reports of {\it oscillatory phenomena in
spicules} appeared in the scientific literature. In particular, it
is anticipated that signatures of the energy transport by MHD waves
through the chromosphere may be detectable in the dynamics of
spicules. A comprehensive review summarizing the current views about
the observed waves and oscillations in spicules, to the best of our
knowledge, is still lacking and such a summary has not been
published yet in the literature. Here we aim to fill this gap.

The goal of this review is to collect the reported observations
about oscillations and waves in spicules, so that an interested
reader could have a general view of the current standing of this
problem. Here, we concentrate only on observed oscillatory and wave
phenomena of spicules and their interpretations. We are not concerned
about the models of spicule generation mechanisms; the interested
reader may find these latter topics in the recent review by \cite{ste00}
or in \cite{dep06}.

Section 2 is a short summary about the general properties of
spicules, Section 3 describes the oscillation events reported so far
for solar limb spicules, Section 4 outlines the views and discussions
about the interpretation of spicule oscillations and Section 5 summarizes
the main results and suggests future directions of research.

\section{General properties of spicules}
\label{sec:gen_prop}

Spicules appear as grass-like, thin and elongated structures in
images of the solar lower atmosphere and they are usually detected
in chromospheric H$\alpha$, D$_3$ and Ca II H lines. These spiky
dynamic jets are propelled upwards (at speeds of about 20 km
s$^{-1}$) from the solar 'surface' (photosphere) into the magnetized
low atmosphere of the Sun. According to early, but still valid
estimates by \cite{wit83} spicules carry a mass flux of about two
orders of magnitude that of the solar wind into the low solar
corona. With diameters close to observational limits ($<$500 km),
spicules have been largely unexplained. The suggestion by
\cite{dep04} and \cite{dep06} of channeling photospheric motion,
i.e. the superposition of solar global oscillations {\it and}
convective turbulence, has opened new avenues in the interpretation
of spicule dynamics (see also {\cite{han06,dep07c,rou07,heg07}}).
The real strength of the observations and the forward modelling by
De Pontieu and Erd\'elyi is that, as opposed to earlier existing
models, they could account {\it simultaneously} for spicule
ubiquity, evolution, energetics and the recently discovered
periodicity (\cite{dep03a,dep03b}) of spicules.

Excellent summaries about the general properties of spicule
(labelled these days as type I spicules) have been presented almost
forty years ago by \cite{bec68,bec72} and we broadly recall these
findings here. Moreover, type II spicules were recently discovered
with Hinode and have very different properties from the classical
spicules (\cite{dep07b}).

\begin{figure}
  \includegraphics[width=0.6\textwidth,angle=90]{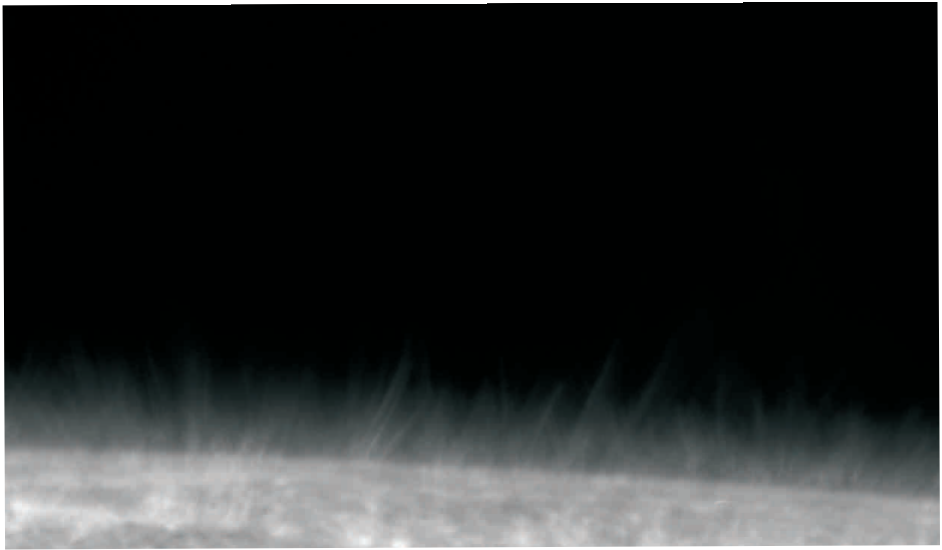}
\caption{High resolution image of spicules at the solar limb in Ca
II H line taken by Solar Optical Telescope (SOT) on board of Hinode
spacecraft (November 22, 2006).}
\label{fig:1}       
\end{figure}

\subsection{Diameter}

\begin{figure}
  \includegraphics[width=0.65\textwidth,angle=90]{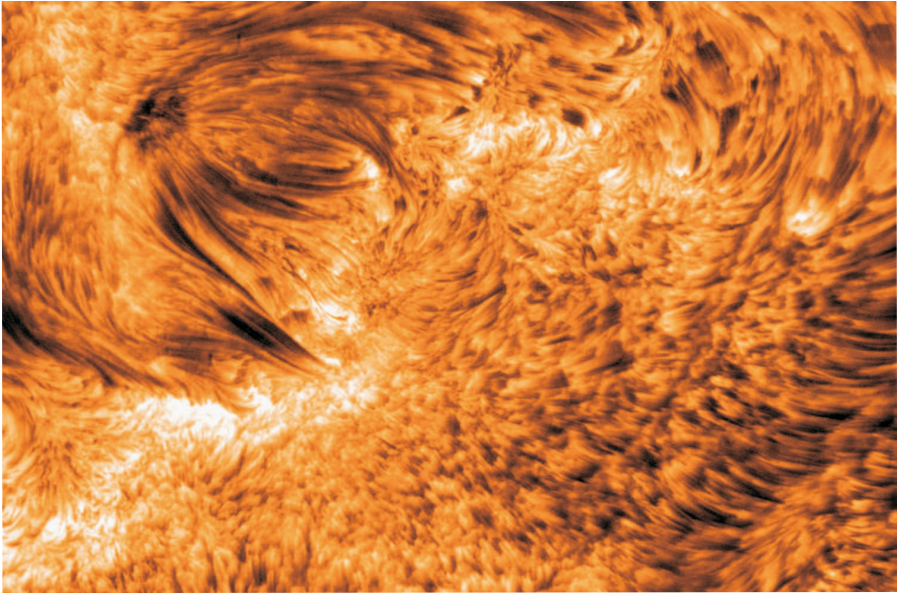}
\caption{High resolution image of spicules on solar disc taken by
Swedish Solar Telescope (SST) in La Palma, adopted from
\cite{dep04}.}
\label{fig:0}       
\end{figure}

Measured range of spicule diameter from ground based observations
was $\sim$ 700-2500 km (\cite{bec68}). The general view was that the
diameter varies from spicule to spicule having the values from 400
km to 1500 km. The spicules seemed to be wider in Ca II H line than
in H$\alpha$ (\cite{bec72}). However, the unprecedentedly high
spatial resolution of Solar Optical Telescope on board of Hinode
spacecraft (0.05 arc sec for Ca II H and 0.08 arc sec for H$\alpha$)
revealed fine structure of spicules. Fig 1. shows this fine
structure of spicules in Ca II H line at the solar limb taken by
Hinode/SOT on November 11, 2006. The type II spicules discovered by
Hinode/SOT have smaller diameters ($\le$ 200 km) in Ca II H line
(\cite{dep07b}), while the diameter of spicules in H$\alpha$ line
seems to be wider $\sim$ 350-400 km (see also the high resolution
image in Fig.~\ref{fig:0} taken by Swedish Solar Telescope, courtesy
\cite{dep04}).

\subsection{Length}

The upper part of spicules continuously fade away with height,
therefore the length is difficult to determine with precision.
Generally, the top of a spicule is defined as the height where the
spicule becomes invisible. The mean length of classical spicules
varies from 5000 to 9000 km in H$\alpha$ (\cite{bec72}) and may
reach to 7000-11000 km heights from the limb when observing by
ground based coronagraphs. On the other hand, the type II spicules
dominate in lower heights: they are tallest in coronal holes
reaching heights of 5000 km or more, while in quiet Sun regions they
reach lengths of order several megameters and they are shorter in
active regions (\cite{dep07b}). Additionally, very long spicules,
called as {\it macrospicules} by \cite{boc75} with typical length of
up to 40 Mm are frequently observed mostly near the polar regions as
reported by e.g.
\cite{pik97,pik98,ban00,par02,yam05,doy05,mad06,osh05,scu09,nis09}.

\subsection{Temperature and Densities}

Spicules have the temperatures and densities typical to the values
of the chromospheric plasmas. Table~\ref{tab:1} summarizes the
typical electron temperatures ($T_e$) and number densities ($n_e$)
of spicule values at different heights above the limb
(\cite{bec68}). Caution has to be exercised as values at 2000 and
10000 km heights are unreliable because of insufficient data.
Typical electron density at the heights, where spicules are
observed, is much lower $\sim$ $10^{9}$  cm$^{-3}$ (\cite{asc04},
Fig 1.19 therein), therefore spicules are much denser than their
surroundings. \cite{mat88} estimated lower temperatures ($\sim$
5000-8000 K) than \cite{bec68}; in general, spicule temperature
seems to be much lower than that of the surrounding coronal plasma.

\begin{table}[ht!]
\caption{Electron temperatures and densities inferred from spicule
emission, after \cite{bec68}}
\label{tab:1}       
\begin{tabular}{lll}
\hline\noalign{\smallskip}
h(km) & $T_e (K)$  & $n_e$ (cm$^{-3}$)  \\
\noalign{\smallskip}\hline\noalign{\smallskip}
2 000 & 17 000 & 22${\times}$10$^{10}$   \\
4 000 & 17 000 & 20${\times}$10$^{10}$  \\
6 000 & 14 000 & 11.5${\times}$10$^{10}$ \\
8 000 & 15 000 & 6.5${\times}$10$^{10}$  \\
10 000 & 15 000 & 3.5${\times}$10$^{10}$   \\

\noalign{\smallskip}\hline
\end{tabular}
\end{table}

\subsection{Life time and motions}

The change of spicule length has been studied by many authors (see
\cite{bec72} and references therein). The general opinion is that
spicules rise upwards with an average speed of 20-25 km/s, reach the
height 9000-10000 km, and then either fade or descend back to the
photosphere with the same speed. The typical life time of classical
spicules is 5-15 mins, but some spicules may live longer or shorter.
On the other hand, the type II spicules from Hinode/SOT have much
shorter life time, about 10-150 s and higher velocities of 50-100
km/s (\cite{dep07b}). Measurements of Doppler shifts in classical
spicule spectra revealed the velocity of 25 km/s, similar to the
apparent speed, therefore it was suggested that the apparent motion
is real. However, it is also possible that observed Doppler shifts
partly correspond to the transverse motions of the spicule axis and
not to the actual movement along the axis. Such transversal motions
can be caused due to the propagation of e.g. waves in spicules.
These periodic perturbations are the subject of our discussion in
the remaining part of the paper.

It must be mentioned here, that some observations show the tilt of
spicule spectra relative to the direction of dispersion, which was
explained as due to the rotation of spicules around their axes
(\cite{pas68,rom75,pis94}). In this regards, it is interesting to
note the recent SOHO/CDS observations, which also suggest the
rotation in macro-spicules, interpreted as a sort of giant solar
tornado (\cite{pik98}).

\section{Oscillations in solar limb spicules}
\label{sec:oscill}

Oscillations in solar limb spicules can be detected either by
imaging or spectroscopic observations. Imaging observations may
reveal the oscillations in spicule intensity and the visual periodic
displacement of their axes. Imaging observations became especially
important after the recently launched Hinode spacecraft. SOT (Solar
Optical Telescope) on board of Hinode gives unprecedented high
spatial resolution images of chromosphere (see Fig. 1). On the other
hand, the spectroscopic observations may give valuable information
about spicules through the variation of line profile. Variations in
Doppler shift of spectral lines can provide information about the
line-of-sight velocity. Through spectral line broadening it is
possible to estimate the non-thermal rotational velocities leading
to the indirect observations of e.g. torsional Alfv\'en waves as
suggested by \cite{erd07}, and reported recently by \cite{jes09} in
the context of a flux tube connecting the photosphere and the
chromosphere. Jess et al. used the technique of analysing
Doppler-shift variations of spectral lines, based on the optically
thick H${\alpha}$ line for which a straightforward interpretation of
linewidth changes and intensity changes in terms of velocity and
density are sometimes very difficult and appropriate caution has to
be exercised, and detected oscillatory phenomena associated with a
large bright point group, located near solar disk centre. Wavelet
analysis reveals full-width half-maximum oscillations with
periodicities ranging from 126 to 700 s originating above the bright
point, with significance levels exceeding 99\%. These oscillations,
2.7 km s$^{-1}$ in amplitude, are coupled with chromospheric
line-of-sight Doppler velocities with an average blue-shift of 23 km
s$^{-1}$. The lack of co-spatial intensity oscillations and
transversal displacements rule out the presence of magneto-acoustic
wave modes. The oscillations are interpreted as a signature of
torsional Alfv\'en waves, produced by a torsional twist of $\pm$25
degrees. A phase shift of 180 degrees across the diameter of the
bright point suggests these Alfv\'en oscillations are induced
globally throughout the entire brightening. The estimated energy
flux associated with this wave mode seems to be sufficient for the
heating of the solar corona, once dissipated. The question
self-evidently emerges: Could spicules guide similar (torsional)
Alfv\'en waves and leak them
to the upper solar atmosphere?\\
Let us return to the possibility of intensity variations of
spicules. Variation of line intensity indicates the propagation of
compressible waves. And finally, the visible displacement of spicule
axis may reveal the transverse waves and oscillations in spicules.
Note that ground based coronagraphs can play an especially important
role in spectroscopic observations. Spatial resolutions of ground
based coronagraphs reach to $\sim$ 1 arcsec, which is less than the
resolution of Hinode/SOT. However, observations on coronagraphs
usually are performed at 5000-10000 km above the surface, where
spicules are less frequent and well separated. Hence, the appearance
of two unresolved spicules inside 1 arcsec is unlikely (note, that
the newly discovered type II spicules do not usually reach these
heights). Therefore, the spectroscopic observations give valuable
information about the dynamics of individual spicule at higher
heights.

In this section we briefly outline almost all oscillation events in solar
limb spicules reported so far in literature. Each observation is
described in a separate subsection. At the end of the section, we
summarize the information gathered about the typical observed periods
in limb spicules.

\subsection{\cite{nik67}}

A very early, to the best of our knowledge the first, modern account of
Doppler shift temporal variation in spicules was reported by \cite{nik67}.
A set of spectrograms of chromospheric spicules in H$\alpha$ line
were obtained with the coronagraph of the Institute of Terrestrial
Magnetism (Russia) on 1 August 1964. The H$\alpha$ profiles and radial
velocities of 11 different spicules were successfully derived from
successive H$\alpha$ line spectra formed at a height of $\sim$6000 km above
the solar limb. The time duration of observations was 8 mins with the
intervals between exposures 10-40 s.
\begin{figure}
  \includegraphics[width=0.5\textwidth]{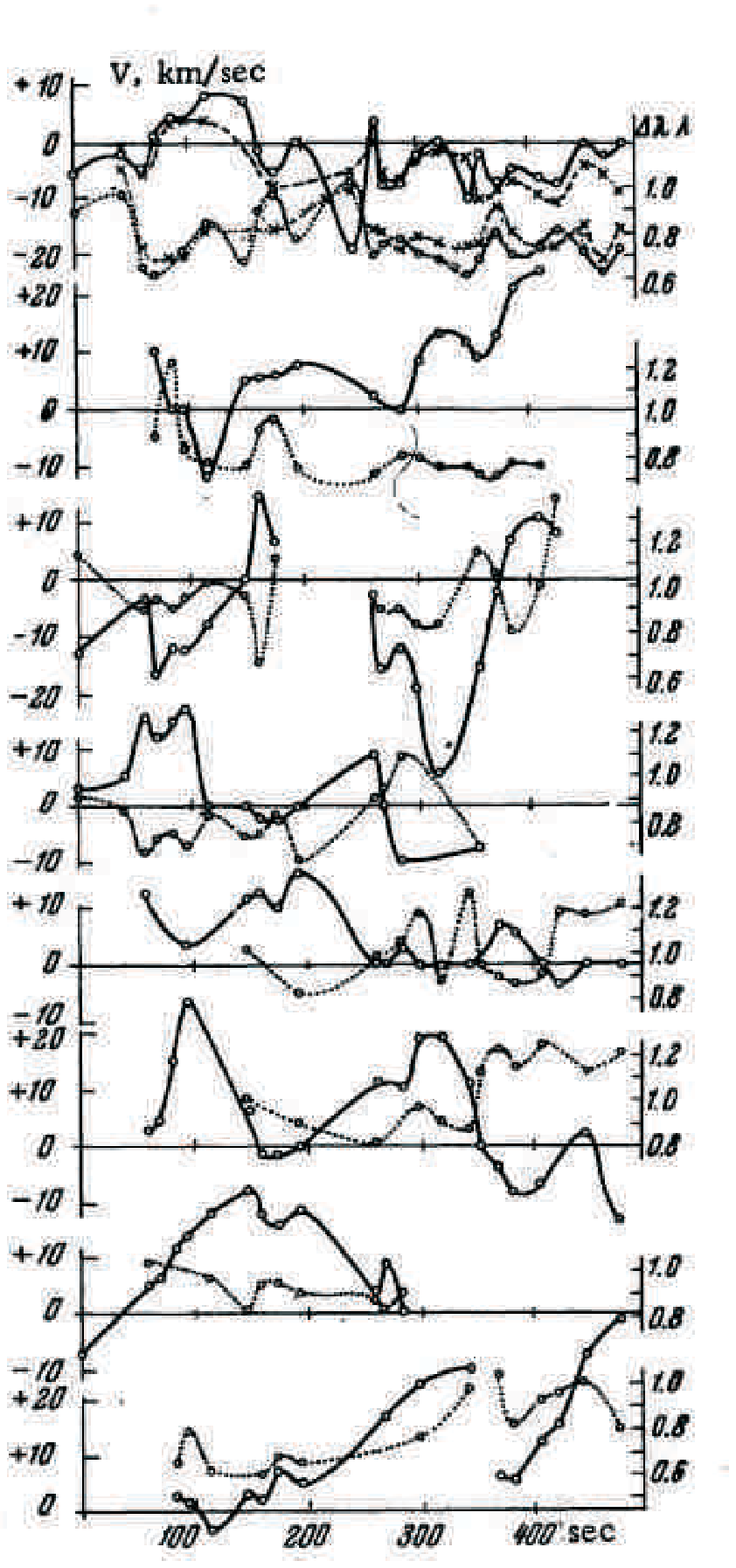}
\caption{Temporal variations in radial velocities (solid curves) and
half widths (dotted curves) of 10 individual H$\alpha$ spicules, adapted
from \cite{nik67}. }
\label{fig:2}       
\end{figure}
Fig.~\ref{fig:2} displays the time evolution of radial velocity and
half width of H$\alpha$ line for 10 different spicules.
Quasi-periodic oscillations are clearly seen. The authors concluded
that the radial velocities vary randomly with time with a mean
period of $\sim$ 1 min. The amplitude of the oscillations are within
10-15 km/s. The half width of H$\alpha$ line profile also tends to
oscillate with a period similar to the mean period of $\sim$ 1 min.
The periodicity is similar to the time scale of type II spicules.
Therefore, this may give the idea that the observed temporal
variations are caused by the type II spicules. However, the type II
spicules usually do not reach the observed heights. Therefore, the
observed oscillations are unlikely to be connected to their
activity.

\subsection{\cite{pas68}}

\cite{pas68} analyzed the high-dispersion spectra of the solar
chromosphere obtained at the Sacramento Peak Observatory in several
spectral regions separately during the summer of 1965. The
observations were carried out simultaneously at two heights in the
solar chromosphere separated by several thousands of kilometers.
\begin{figure}[ht!]
\includegraphics[width=0.7\textwidth]{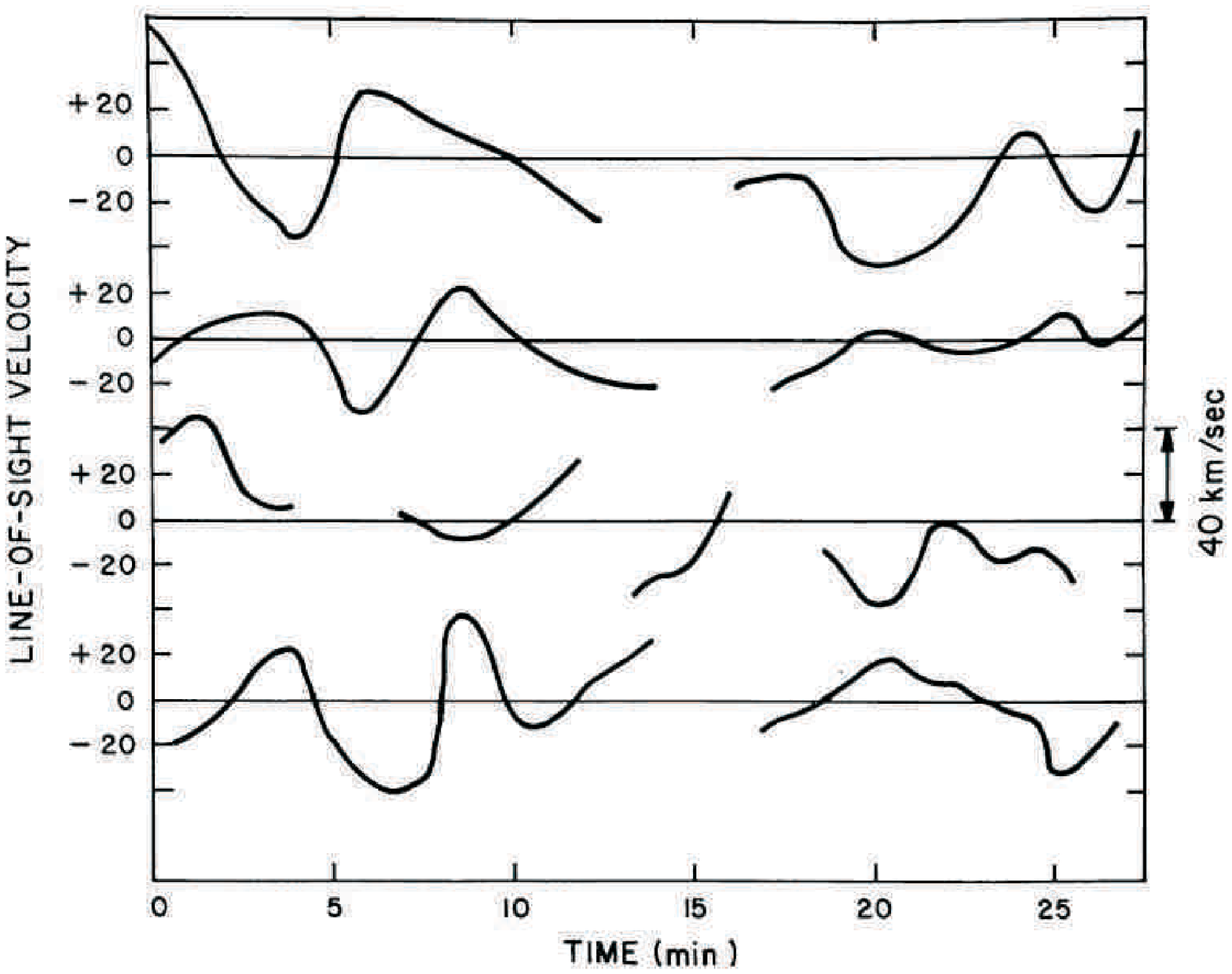}
\caption{Temporal variation of Doppler velocities in H-line of Ca II
from \cite{pas68}.}
\label{fig:3}       
\end{figure}

The time series of Doppler velocities in H-line of ionized Calcium,
Ca II, for 10 different spicules at the height of 5000 km are shown
in Fig.~\ref{fig:3}. The exposure time for H-line was 13 s and the
spatial resolution was less than 2 arc sec. \cite{pas68} were
searching for the sign reversal of Doppler velocities in order to
determine the rising/falling stages of spicule evolution. Indeed,
some features show the sign reversal, but the common property is the
clear quasi-periodic temporal variation of Doppler velocities with
periods of 3-7 min. The amplitudes of oscillations are rather high,
though still being within the range of 10-20 km/s. \cite{pas68}
interpreted the detected temporal variation as motion along the
spicule axis, but transverse oscillations also cannot be ruled out.

\subsection{\cite{wea70}}

Observations have been carried out with the Mount Wilson Solar Tower
Telescope during the period of 10 September - 13 October, 1967. Time
sequences of H${\alpha}$ spectra with time lapse rates of 5 to 15 s
have been obtained corresponding to height 5000-6000 km height above
the solar limb.

The author reported that, both Doppler velocity and horizontal
motion of spicules as a whole have significant input into spicule
dynamics. In at least two cases, the author found that the combined
motion indicate movement of a gas in an arc of a horizontal circle,
firstly towards the observer, followed by sideways, finally away. Weart
concluded that only true transverse motion could explain the observed
pattern of motion.

The power spectrum of temporal variations resembled the familiar
1/frequency curve, typical to many types of random motions.
Substantial power was found to be concentrated at periods of 1, 2.5 and
10 minutes. However, no statistically significant peaks were
observed. Therefore, it was concluded that spicules move
horizontally at random.

\subsection{\cite{nik71}}

Detailed spectroscopic observations were carried out on 3 April 1969
with the 53 cm Lyot coronagraph mounted at the High Altitude
Astronomical Station near Kislovodsk (Russia). 38 H${\alpha}$
spectrograms of the chromosphere were obtained during about a 21
minute observing campaign at the height of 4200 km. The time
interval between successive frames varied from 14 to 100 s being on
the average about 30 s. The spatial resolution of observations was
$\sim$ 1 arcsec.
\begin{figure}[ht!]
  \includegraphics[width=0.5\textwidth]{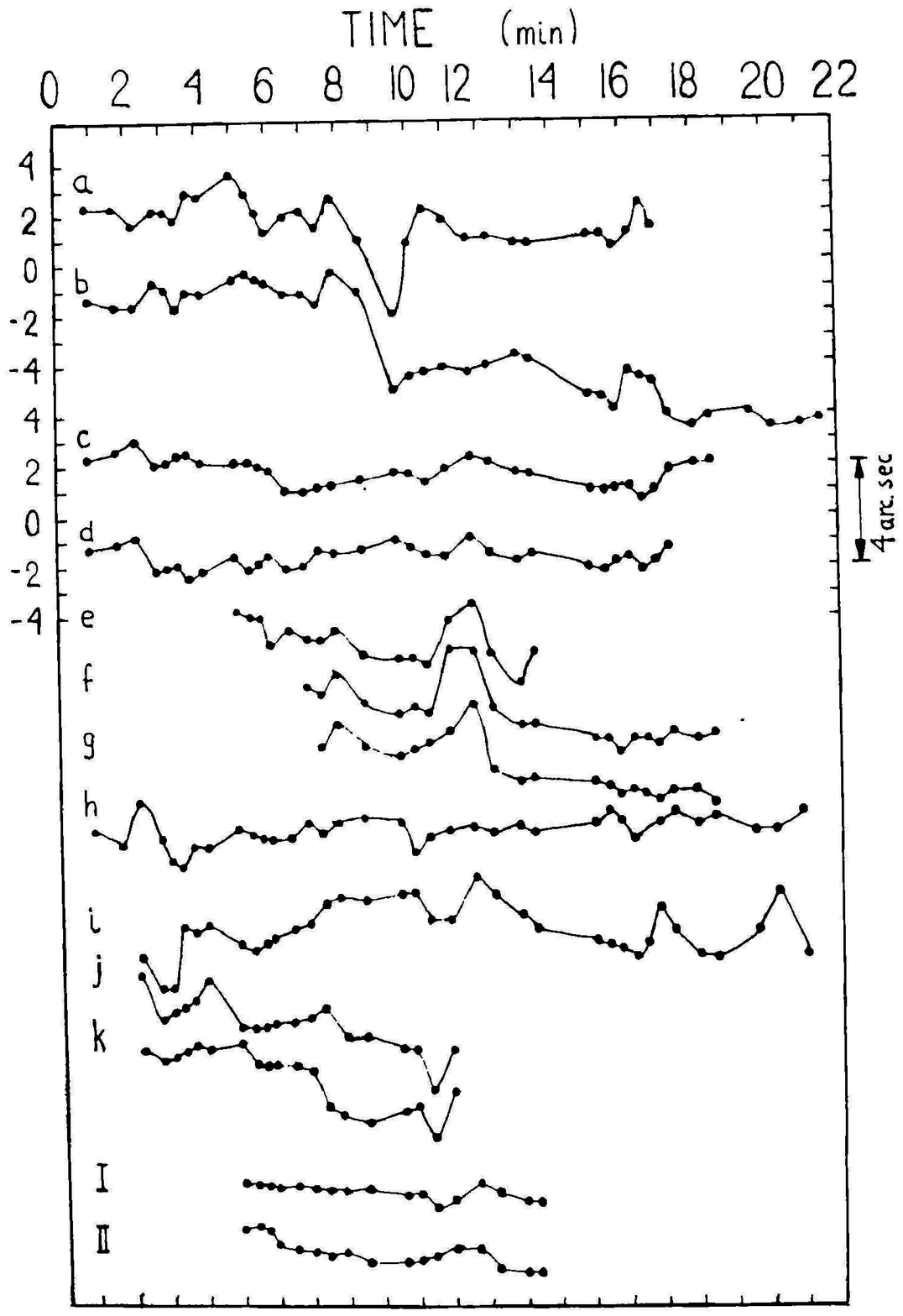}
    \includegraphics[width=0.4\textwidth,angle=90]{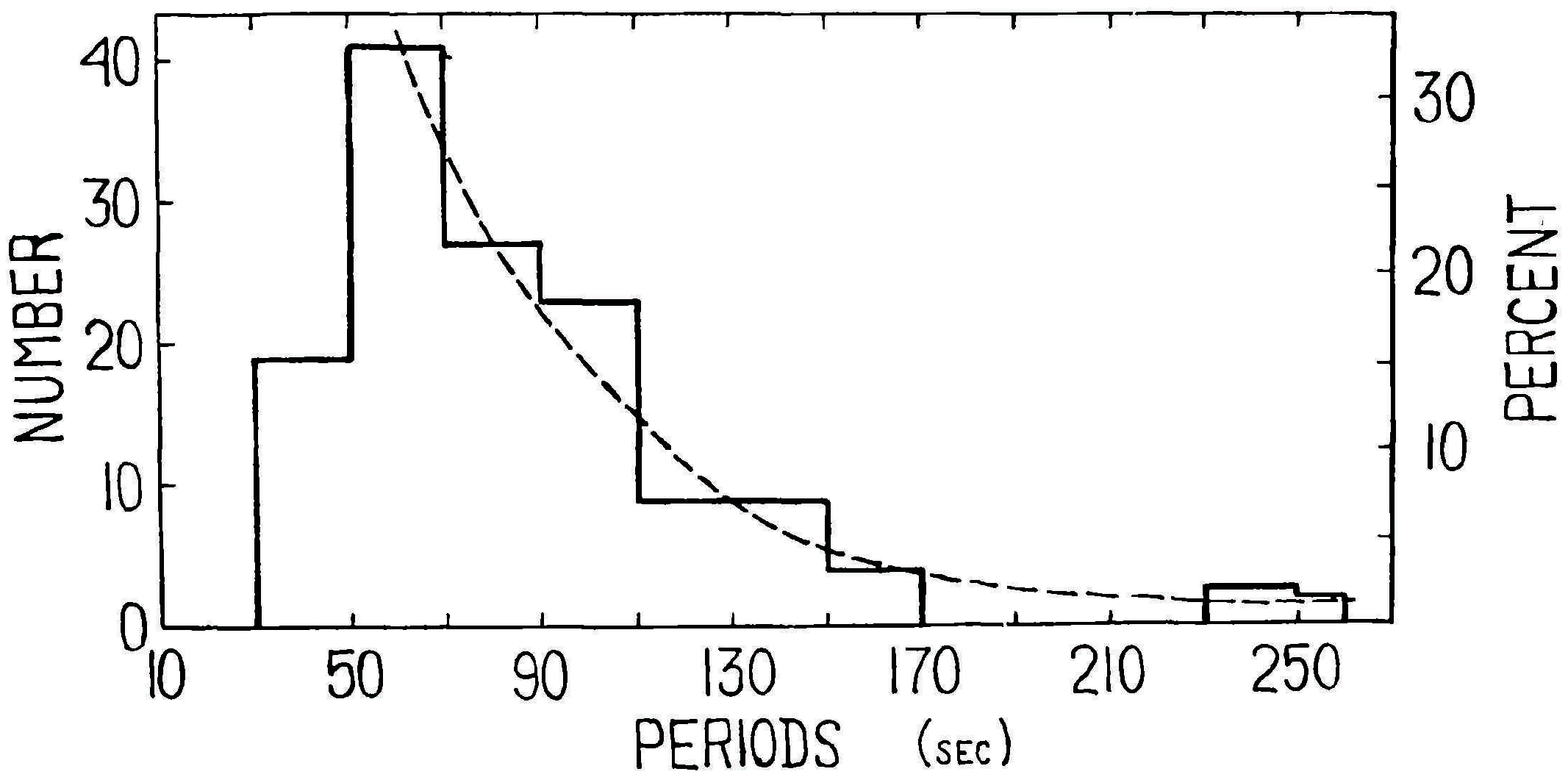}
\caption{Left: temporal variation of spicule positions along the
solar limb, from \cite{nik71}. Points denote the different positions
of spicules during time series. I and II are the "bench-mark"
spicules. Right: distribution of periods of spicule oscillations
along the solar limb. }
\label{fig:4}       
\end{figure}
Two comparatively stable spicules, present in almost all frames,
were chosen as reference ones. Then, the variation of other spicules
along the limb with respect to these "bench-mark" spicules were
determined. Fig.~\ref{fig:4} (left panel) shows the position of
several spicules vs time with respect to the "bench-mark" spicules.
There is evidence of oscillations of spicule position along the
limb.

The distribution of periods of spicule oscillations along the limb
is shown on the right panel of Fig.~\ref{fig:4}. The most probable
period lies between 50-70 s and the authors concluded that spicules
undergo transversal oscillations with a period of $\sim$ 1 min. The
amplitude of oscillations was estimated to be about 10-15 km
s$^{-1}$. The observations have been performed at the height, where
the type II spicules may reach, therefore it is possible that
observed temporal variations here are connected to their activity.

\subsection{\cite{kul78}}

The observational material described in the previous subsection
was re-analyzed later by \cite{kul78} in order to search for a longer
periodicity. The authors also searched for signatures of possible
oscillations in Doppler velocity, line width and intensity, respectively.
\begin{figure}[h!]
  \includegraphics[width=0.5\textwidth,angle=90]{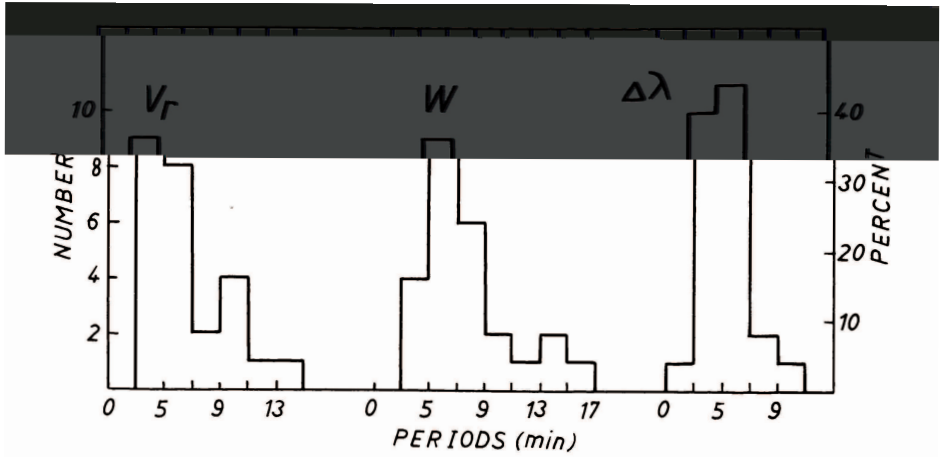}
\caption{Distribution of periods of Doppler velocity $V_r$,
intensity $W$ and full-width at half-maximum $\Delta \lambda$,
adapted from \cite{kul78}.}
\label{fig:5}       
\end{figure}
Fig.~\ref{fig:5} shows the distribution of the spicule number vs the
observed periods of oscillation in line-of-sight velocity, line width and
intensity. About 70\% of the observed periods of Doppler shift
oscillations are within 3-7 min. The same per cent of observed
periods lies within 4-9 min in the spicule intensity and 80\% of
observed periods of line width oscillations are within 3-7 min.

\subsection{\cite{gad82}}

Observations were also carried out with the 53 cm Lyot coronagraph
at Shemakha Astrophysical Observatory resulting in H${\alpha}$ time
series corresponding to a height of 4 Mm above the solar limb. 26
spectrograms have been taken over an 8-min interval which gives
$\sim$ 20 s between consecutive frames. A total number of 15
spicules were investigated in details. Gadzhiev and Nikolsky
analysed variations in Doppler velocity as well as in the tangential
velocity, i.e. reflecting the visible displacement of spicule axes
along the solar limb.

\begin{figure}
\includegraphics[width=0.5\textwidth]{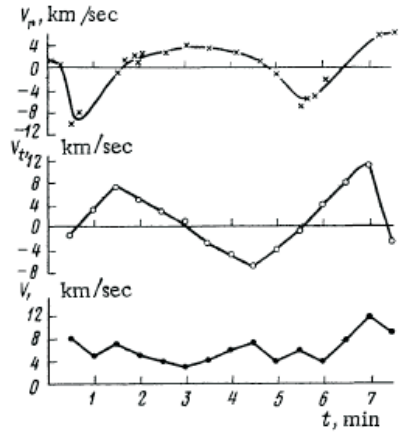}
\caption{Time variation of the radial and tangential velocities
$V_r$, $V_t$ and the modulus $V$ of the velocity vector for a
spicule, adapted from \cite{gad82}.}
\label{fig:6}       
\end{figure}

The authors found that the spicules oscillate with typical periods of
3-6 mins, both in line-of-sight and tangential directions. Fig.~\ref{fig:6}
shows the time variation of line-of-sight and tangential velocities in one of
the spicules. The periodicity in both velocity components is clearly visible.
Gadzhiev and Nikolsky also constructed the trajectories of spicule motion
by putting together both velocity components. They concluded that
spicules undergo a cyclic motion as a whole on an ellipse with an average
period of 4 mins. The average amplitude of this cyclic motion was 11 km s$^{-1}$.

\subsection{\cite{kul83}}

Another spectroscopic observations were carried out with the 53 cm
Lyot coronagraph mounted at the Abastumani Astrophysical Observatory
(Georgia). A 22 minutes long time sequence (the time cadence is not
known), taken in the H${\alpha}$ line, was analyzed for a total
number of 25 spicules. The statistically significant period of
oscillations in intensity, line width and line-of-sight velocity was
found to be $\sim$ 5 min.

\subsection{\cite{has84}}

Observations were also carried out with the Vacuum Tower Telescope of the
National Solar Observatory (USA) in the autumn of 1982. A set of
H${\alpha}$ spectra corresponding to five slit positions above the
solar limb were recorded every 8 s in order to investigate the temporal
variations of spicules. The spatial resolution of these observations was
better than 2 arcsec.

Hasan and Keil detected the temporal variations of the line-of-sight
velocity at two different heights for two spicules. The fine time
resolution allowed them to discern small amplitude fluctuations with
periods of about 2-3 mins.

\subsection{\cite{pap94}}

Observations were carried out with the 53 Lyot coronagraph of Sayan
Observatory located near Irkutsk (Russia). The spectroscopic time
series in different spectral lines varied from several minutes to
hours with an excellent temporal resolution of 10-20~s. The spatial
resolution of observations was better than 1 arcsec. The spectra
were simultaneously registered at three different heights (5000
-8000 km above the limb) above the limb with a three-level image
slicer.
\begin{figure}[ht!]
  \includegraphics[width=0.6\textwidth]{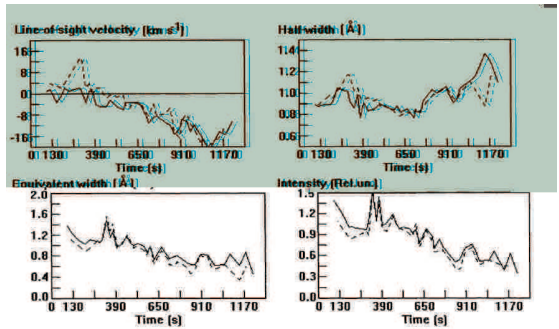}
\caption{Temporal variations of H${\alpha}$ line profile parameters
and the line-of-sight velocity of spicules at different heights, adapted from
\cite{pap94}. The solid line corresponds to the height of 5 Mm and
the dashed line to 8 Mm above the solar limb, respectively.}
\label{fig:7}       
\end{figure}

Temporal variations of H${\alpha}$ line profile parameters and
line-of-sight velocity for one of spicules at two different heights
are shown on Fig.\ref{fig:7}. The quasi periodic fluctuations are clearly
seen. Papushev and Salakhutdinov found that the oscillation periods lay
between 80-120 sec.

\subsection{\cite{xia05}}

Xia et al. analyzed the time series of EUV spicules in two polar
coronal holes obtained by the SUMER (Solar Ultraviolet Measurements
of Emitted Radiation) camera on-board the SOHO (SOlar and Heliospheric
Observatory) spacecraft. The spatial resolution of the observations was 1
arcsec and the exposure time for different data sets varied as 15, 30 and 60
s.
\begin{figure}[ht!]
  \includegraphics[width=0.7\textwidth,angle=90]{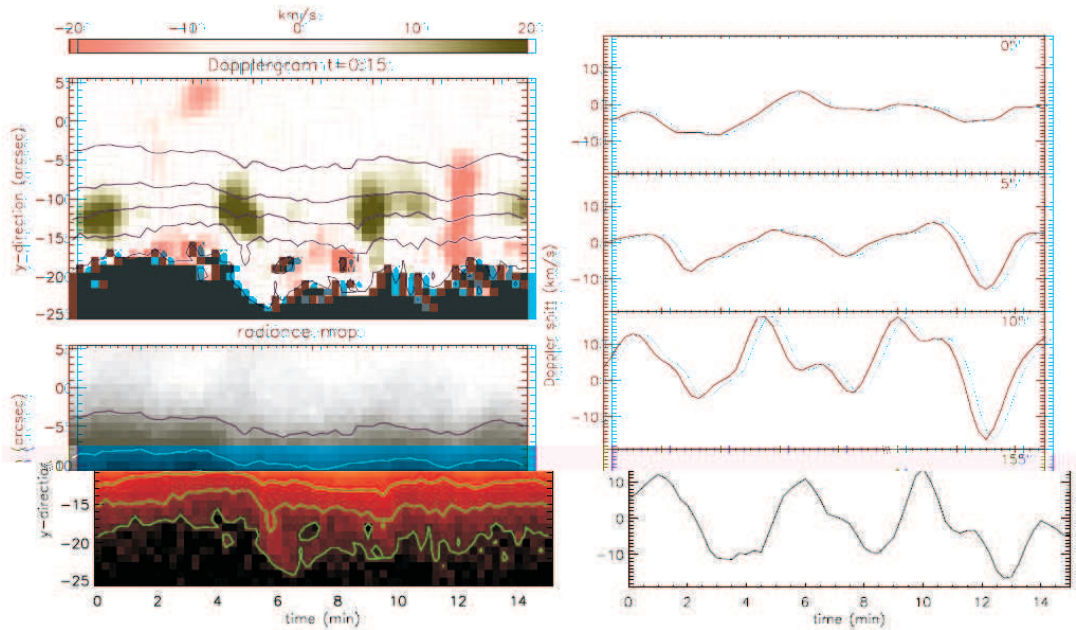}
\caption{Time series Dopplergram and radiance map for the C III 977
\AA\ line showing levels of radiance in logarithmic scale (adapted
from \cite{xia05}). The right panel shows the relative Doppler shift
at four location above the limb. }
\label{fig:8}       
\end{figure}
Fig.~\ref{fig:8} shows Dopplergrams and radiance map for the C~III
977 \AA \ line (left panel). The right panel shows the relative
Doppler shifts at four different locations above the solar limb. The
Doppler velocity and radiance indicate evidence of $\sim$5-min
oscillations.

\subsection{\cite{kuk06}}

Observations were carried out on 26 September 1981 with the 53 cm
Lyot coronagraph of the Abastumani Astrophysical Observatory (the
instrumental spectral resolution and dispersion in H${\alpha}$ are
0.04 {\AA} and 1 {\AA}/mm correspondingly) at the solar limb. The
scanning of height series began at the height of 3800 km measured
from the photosphere, and continued upwards (\cite{khu86}). The
chromospheric H${\alpha}$ line was used again to observe solar limb
spicules at 8 different heights. The distance between neighbouring
heights was 1$^{\prime\prime}$ (which was the spatial resolution of
observations), thus the distance of $\sim$3800-8700 km above the
photosphere was covered. The exposure time was 0.4 s at four lower
heights and 0.8 s at higher ones. The total time duration of each
height series was 7 s. Consecutive height series began immediately,
once a sequence was completed.
\begin{figure}[ht!]
  \includegraphics[width=0.6\textwidth]{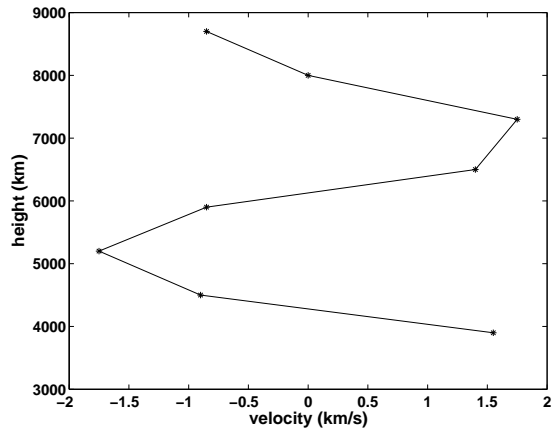}
\caption{The Doppler velocity spatial distributions for one of the
height series from \cite{kuk06}. The marked dots indicate the observed
heights.}
\label{fig:9}       
\end{figure}
\cite{kuk06} analyzed the spatial distribution of Doppler velocities
in selected H${\alpha}$ height series. Nearly 20$\%$ of the measured
height series showed a periodic spatial distributions in the Doppler
velocities. A typical Doppler velocity spatial distributions for one
of the height series is shown in Fig.~\ref{fig:9}, which shows a
periodic behavior. The authors suggested that the spatial
distribution was caused by transverse kink waves. The wavelength was
estimated to be ${\sim}$3.5 Mm. The period of waves was estimated to
be in the range of 35-70 s.

\subsection{\cite{zaq07}}

Zaqarashvili et al. analysed the same observational data obtained by the 53
cm Lyot coronagraph of the Abastumani Astrophysical Observatory by
\cite{khu86}. Using the height series, they constructed continuous
time series of H${\alpha}$ spectra with an interval of $\sim$7-8 s
between consecutive measurements at each height. The time series
cover almost the entire lifespan (from 7 to 15 mins) of several spicules.
\begin{figure}[ht!]
  \includegraphics[width=0.6\textwidth]{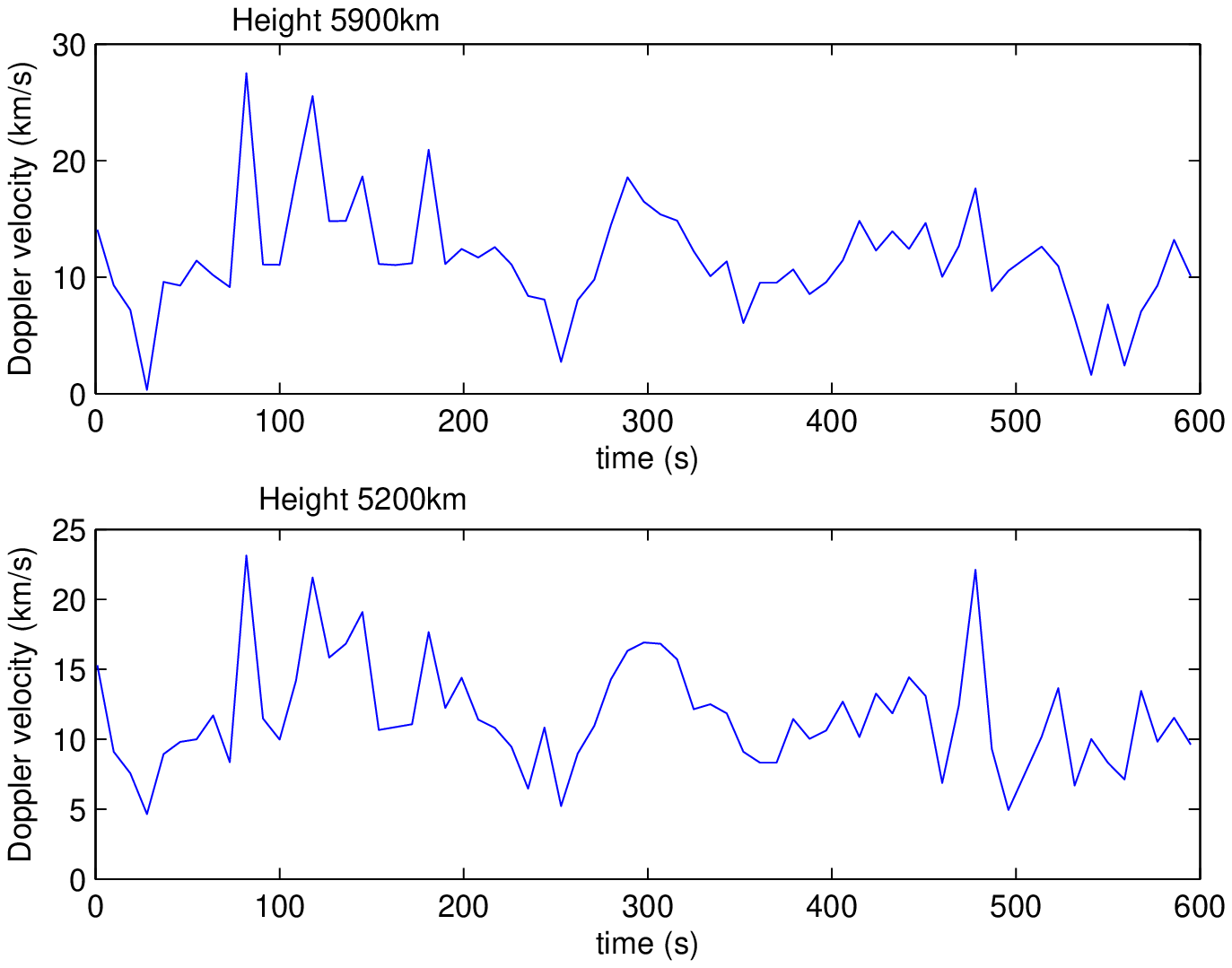}
  \includegraphics[width=0.65\textwidth]{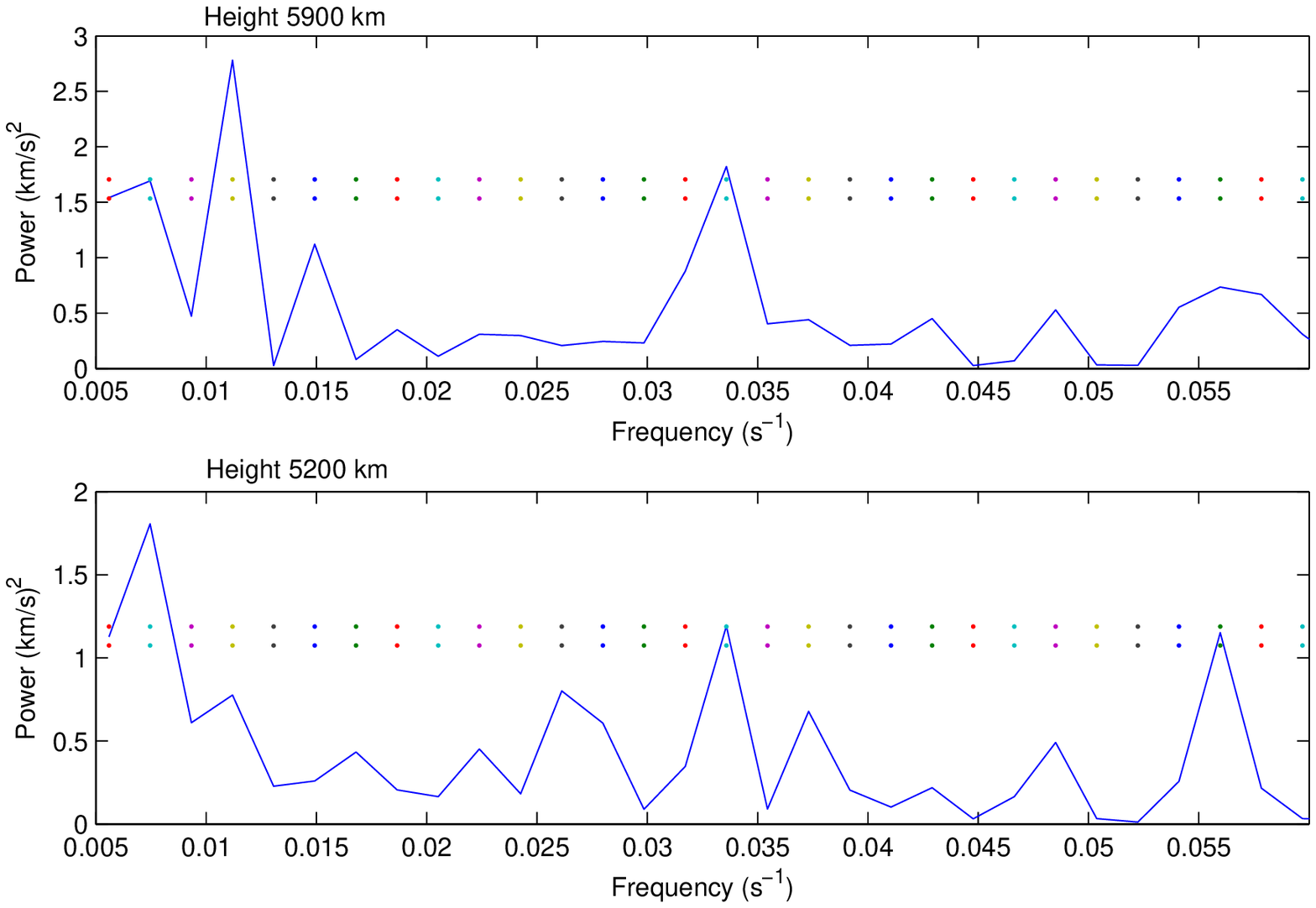}
\caption{Left: Doppler velocity time series at the heights of 5200
and 5900 km in one of the spicules, adapted from \cite{zaq07}. The
time interval between consecutive measurements is $\sim$8 s. Right:
power spectra of Doppler velocity oscillations from the time series.
The dotted lines in both plots show 95.5$\%$ and 98$\%$ confidence
levels, respectively. There is the clear evidence of oscillations
with periods of 180 and 30 s at both heights.}
\label{fig:10}       
\end{figure}
Figure~\ref{fig:10} shows the Doppler velocity time series at two
different heights above the solar limb in one of the spicules (left
panel). The time series show the evidence of quasi-periodic
oscillations in line-of-sight velocity. The power spectra resulted
from Discrete Fourier Transform (DFT) analyses of the time series
are presented in the right panel. The most pronounced periods at
both heights are 180 and 30 s.  The oscillation with the period of
90 s is also seen but preferably at higher heights (note the small
peak at the lower height as well).
\begin{figure}
  \includegraphics[width=0.6\textwidth]{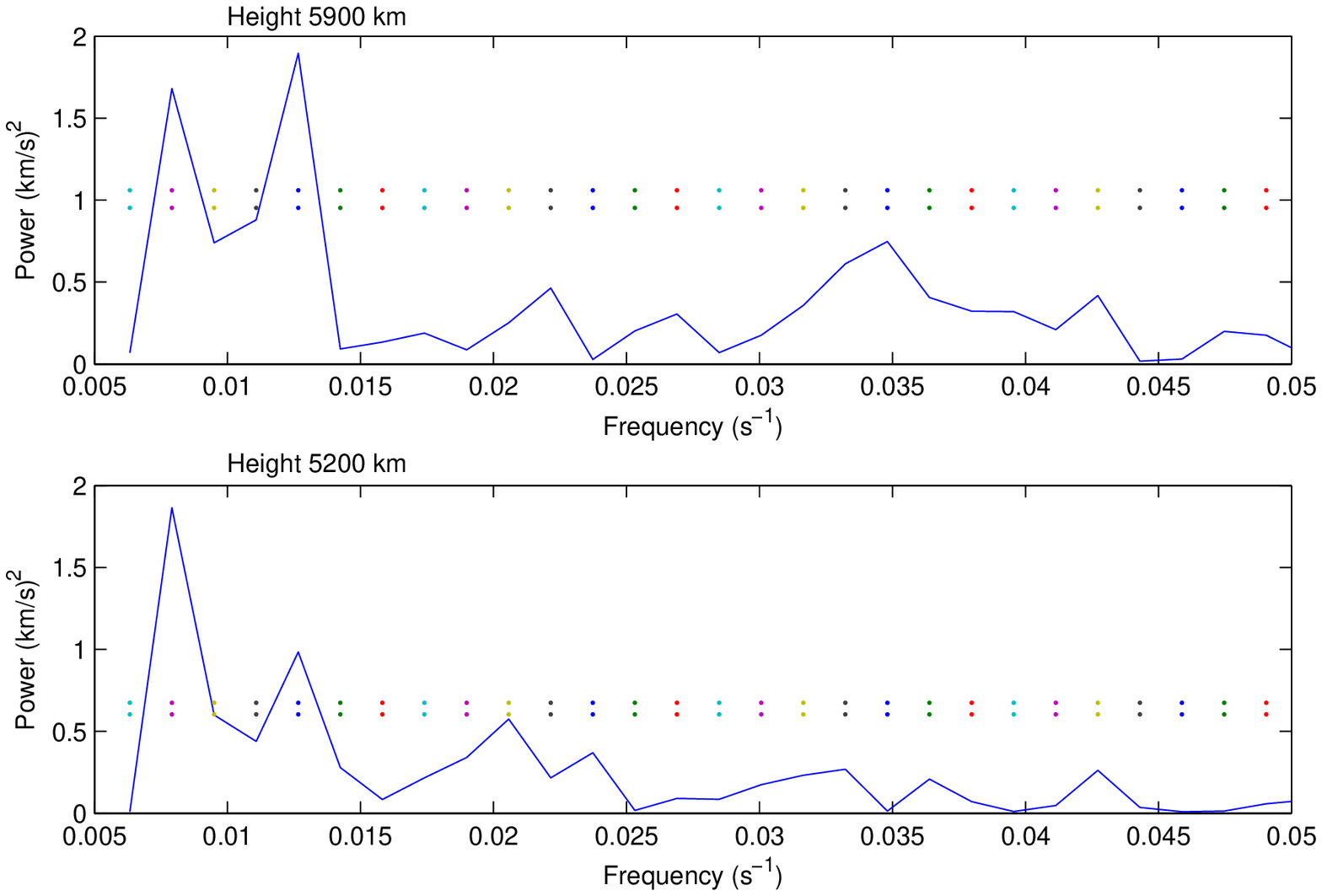}
  \includegraphics[width=0.6\textwidth]{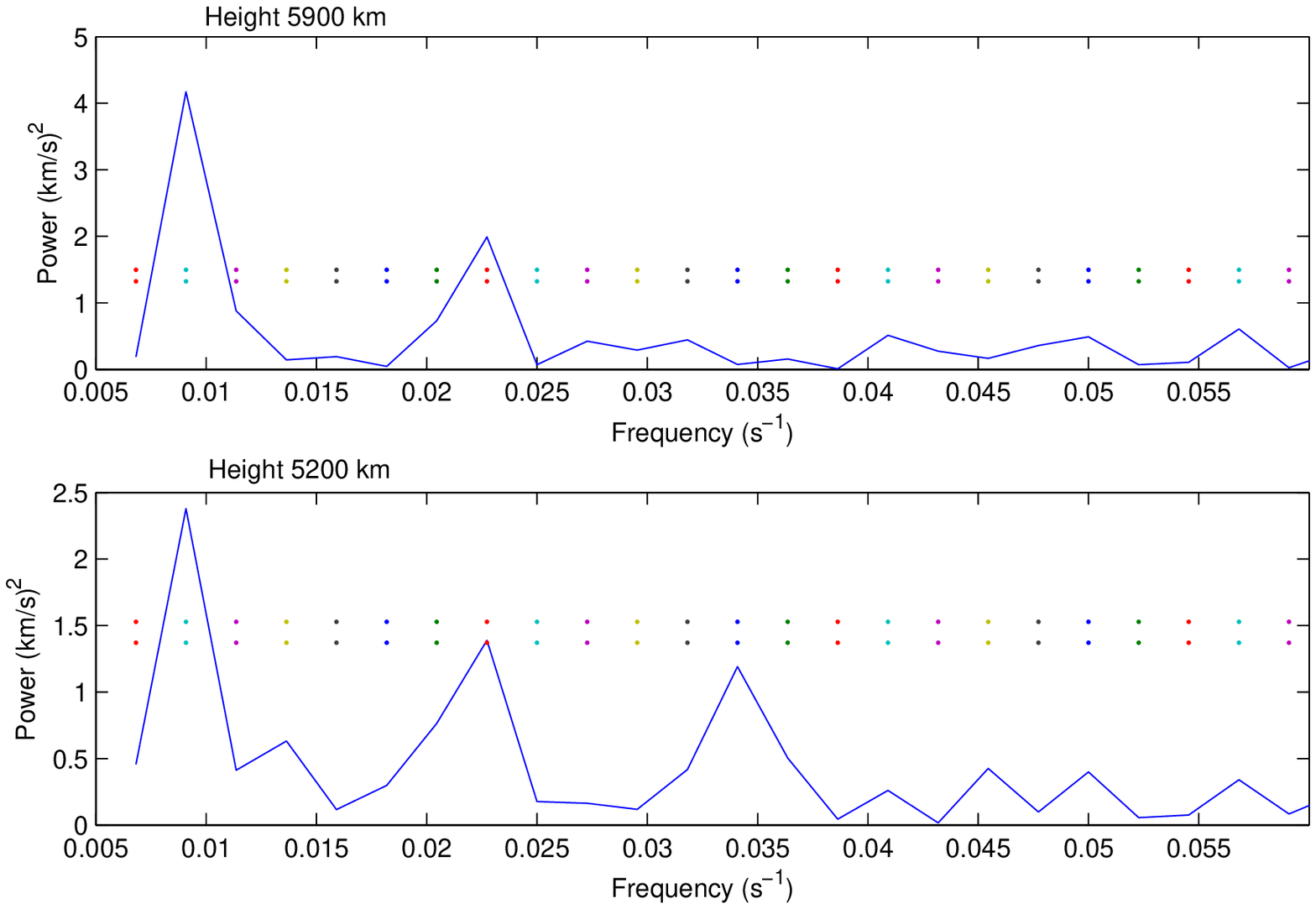}
\caption{Power spectra of Doppler velocity oscillations in another
two spicules at the heights of 5200 and 5900 km (\cite{zaq07}). The
dotted lines in both plots show 95.5$\%$ and 98$\%$ confidence
levels. }
\label{fig:11}       
\end{figure}

The power spectra of Doppler velocity oscillations in two other
spicules at the heights of 5200 km (lower panels) and 5900 km (upper
panels) are plotted on Figure~\ref{fig:11}. One of the spicules
(left panels) shows the two clear oscillation periods of 120 and 80 s
at both heights. Both periods are above the 98$\%$ confidence level.
Another spicule undergoes oscillations with periods $\sim$110 and
$\sim$40 s, respectively.

\begin{figure}
  \includegraphics[width=0.9\textwidth]{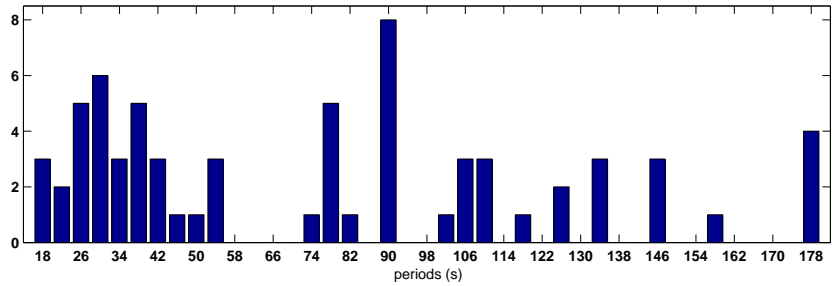}
\caption{Histogram of all oscillation periods that are above
95.5$\%$ confidence level, adapted from  \cite{zaq07}. The horizontal axis
shows the oscillation periods in seconds, while the vertical axis
shows the number of corresponding periods.}
\label{fig:12}       
\end{figure}

Zaqarashvili et al. also presented the results of DFT for 32 different
time series as a histogram of all the oscillation periods above the
95.5$\%$ confidence level (see Figure~\ref{fig:12}). Almost half of the
oscillatory periods are located in the period range of 18-55 s. Another
interesting range of oscillatory periods is at 75-110 s, with a clear
peak at the period of 90 s. Note that there is a further interesting
period peak at 178 s as well, which is interpreted as a clear evidence
of the well-known 3 min oscillations ubiquitous in the lower solar atmosphere.

\subsection{\cite{dep07a}}

These authors analysed the time series of Ca II H-line images taken
with SOT (Solar Optical Telescope) onboard the recently launched
Hinode satellite. SOT has high spatial (0.2 arcsec) and temporal (5
s) resolutions, and is capable of providing extremely useful
information about the dynamics of the lower atmosphere.

\begin{figure}
  \includegraphics[width=0.8\textwidth]{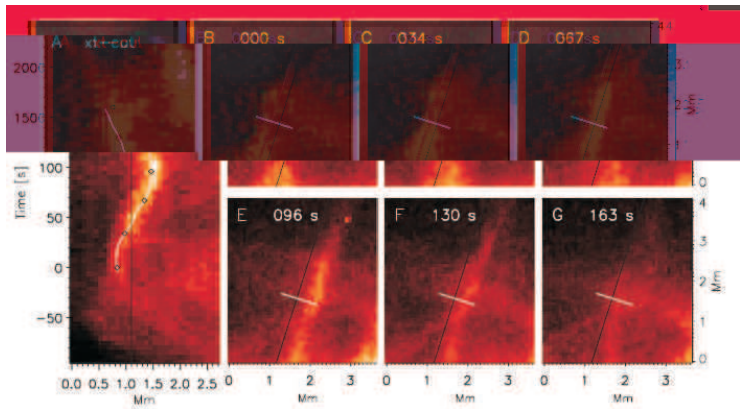}
\caption{An example of the transversal motion of a spicule obtained with Hinode/SOT
by \cite{dep07a}. Panel A shows the intensity as a function of time along the spatial
cut indicated by a white line on panels B-G. Panels B-G represent the time sequence of Ca II H
images.  }
\label{fig:13}       
\end{figure}

\cite{dep07a} found that many of the chromospheric spicules (both,
type I and II) undergo substantial transverse displacements, with
amplitudes of the order of 500 to 1000 km during their relatively
short lifetime of 10 to 300 s. They also reported that some
longer-lived spicules undergo periodic motions in a direction
perpendicular to their axes. Fig.~\ref{fig:13} shows an example of
transverse displacement of a spicule with a period of $\sim$3 mins.
\begin{figure}
  \includegraphics[width=0.9\textwidth]{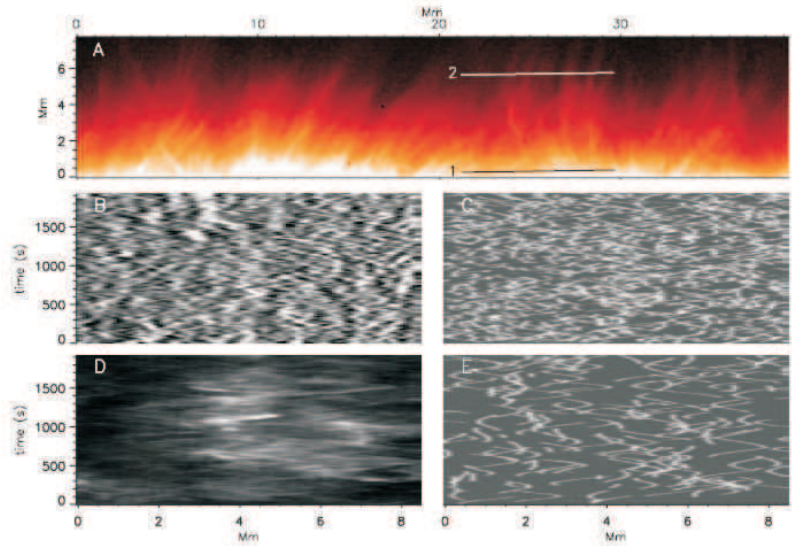}
\caption{Transverse motion of many spicules, from \cite{dep07a}. Panel A
shows an image of the Hinode/SOT Ca II chromospheric line. Panels B and
D are time-distance plots along the cuts labeled by 1 and 2 on panel A.
Panels C and E are similar cuts but reproduced by Monte-Carlo numerical
simulations of spicule motions.
}
\label{fig:14}       
\end{figure}
In general, \cite{dep07a} found it difficult to determine the
periods of transverse motions due to the short life time of
spicules. Fig.~\ref{fig:14} shows the fine structure of spicules in
Ca~II H-line and the time-distance plot along the cut parallel to
the solar surface. The plots reveal the predominant linear motion of
spicules rather than oscillatory ones. However, comparing these
excellent quality observations to their Monte-Carlo simulations of
numerically modelling swaying spicule motion, they suggested that
the most expected periods of transverse oscillations should lay
between 100 and 500 s, which they interpreted as signature of
Alfv\'en waves.

\subsection{Summary of observed oscillatory phenomena}
\label{ssec:sum_obs}

All the observations allow us to make a short summary of oscillatory
phenomena detected in limb spicules. The results are gathered in
Table~\ref{tab:2}.
\begin{table}[ht!]
\caption{Summary of observed oscillatory periods in solar limb spicules}
\label{tab:2}       
\begin{tabular}{llllll}
\hline\noalign{\smallskip}
 & Doppler & displacement & intensity & speed & line\\
\noalign{\smallskip}\hline\noalign{\smallskip}
\cite{nik67} & 1 min & 1 min & & & H${\alpha}$  \\
\cite{pas68} & 3-7 min & & & $>$ 90 km/s & Ca II \\
\cite{wea70} & random & random & & & H${\alpha}$ \\
\cite{nik71} &  & 50-70 s & & & H${\alpha}$  \\
\cite{kul78} & 3-7 min &  & 3-7 min & & H${\alpha}$  \\
\cite{gad82} & 3-6 min & 3-6 min & & & H${\alpha}$  \\
\cite{kul83} & 5-min & & 5-min & & H${\alpha}$  \\
\cite{has84} & 2-3 min &  & & $>$ 300 km/s & H${\alpha}$  \\
\cite{pap94} & 80-120 s & 80-120 & 80-120 & $>$ 300 km/s & H${\alpha}$ \\
\cite{xia05} & 5-min & & 5-min & & EUV\\
\cite{kuk06} & 35-70 s & & & $\sim$ 80 km/s & H${\alpha}$  \\
\cite{zaq07} & 30-110 s & & & $\sim$ 110 km/s & H${\alpha}$  \\
\cite{dep07a} & & 100-500 s & & 50-200 km/s & Ca II \\

\noalign{\smallskip}\hline
\end{tabular}
\end{table}
Oscillations are more frequently observed in the Doppler velocity,
which points towards transversal motions as the observations are
performed at the solar limb. The longitudinal velocity component may
also take a part in shaping the Doppler velocity oscillations as
spicules are generally tilted away from the vertical. However, the
transverse component seems to be the more important and determinant
component in these oscillations as the visible displacement along
the limb is also frequently reported. The observed periods can be
formally divided into two groups: those with shorter periods
($<$2-min) and those with longer periods $\geq$ 2-min. The most
frequently observed oscillations are within the period ranges of
$3-7$ min and $50-110$ s. The two groups of oscillations are
possibly caused by different physical mechanisms, but this issue
needs further studies before one can conclude. Intensity
oscillations are observed mostly with $\sim$5-min period, which may
indicate their connection to the global photospheric 5-min
oscillations (\cite{dep03a,dep03b,dep04,dep06}). However, $\sim$
5-min oscillations in Doppler velocity and visible displacement can
be caused by transverse waves (kink or Alfv\'en). Oscillations with
other periods seem to be connected mostly with transverse
displacement of spicule axes, which can be caused by kink or
Alfv\'en waves. These possibilities are discussed later in details.

In order to infer more precise information about the oscillatory
motions in the lower solar atmosphere, in particular in
chromospheric spicules, it is of vital importance to determine
whether the oscillations are caused by propagating or standing wave
patterns. Unfortunately, this is a very challenging task from an
observational perspective, and it is a difficult task from a
theorist's point of view as well. To the best of our knowledge, very
little is known about these mechanisms applicable to the highly
stratified spicule geometry in the literature. \cite{erdetal07} give
an insight into the theoretical and numerical difficulties of
studying waves and oscillations in the highly stratified lower solar
atmosphere. In spite of these obstacles, some conclusions still can
be drawn and we discuss them in the next section.

\section{Propagation}
\label{sec:prop}


The propagation of disturbances along spicules can be deduced when
observations are performed at least at two different heights. Then
the propagation speed can be estimated from the phase difference
between oscillations at different heights (see Table II).

Several authors reported the possible propagation speeds of
disturbances in spicules. \cite{pas68} found that velocity changes
occur simultaneously, to within 20~s, at two distinct heights
separated by 1800 km. They concluded that the propagation velocities
should be more than 90 km s$^{-1}$. \cite{has84} suggested the
propagation of signals from lower to higher heights with an
estimated velocity of more than 300 km s$^{-1}$. \cite{pap94}
studied the phase delays of fluctuations at different heights and
also concluded that the propagation speeds should exceed 300 km
s$^{-1}$. However, it must be noted that a standing oscillation
pattern may also be responsible to explain these phenomena. Indeed,
Hinode/SOT observations (\cite{dep07a}) show some evidence of upward
and downward propagating waves, with some partially standing waves
being observed. \cite{dep07a} estimated wave phase speed in the
range of 50-200 km/s.

\begin{figure}[h!]
  \includegraphics[width=0.6\textwidth]{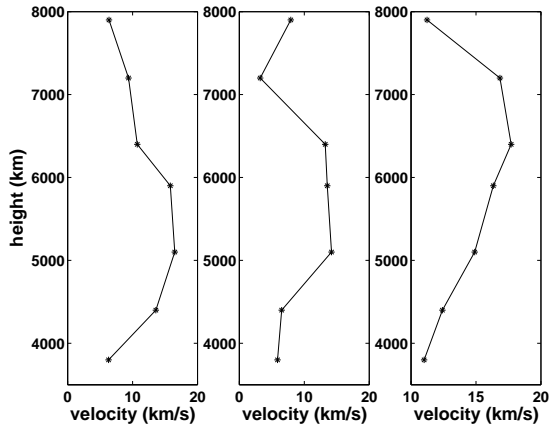}
\caption{Three consecutive height series of Doppler velocities in one
of the spicules, adapted from \cite{kuk06}. The time difference between the
consecutive plots is $\sim$8 s. The maximum of Doppler velocity
moves up in consecutive height series, which most probably indicates
wave propagation.}
\label{fig:15}       
\end{figure}

Detailed reports about wave propagation were presented by e.g. \cite{kuk06}
and \cite{zaq07} through analyzing the consecutive height series.
\cite{kuk06} presented three consecutive height series of Doppler
velocities in a spicule, which show that the maximum of the Doppler
velocity moves up in time (see Fig.~\ref{fig:15}).
The authors suggested that this may indicate a wave phase
propagation. The phase is displaced at $\sim$1500 km in about 18 s
giving the phase speed of $\sim$80~km~s$^{-1}$, very comparable to
the expected kink or Alfv\'en speed at these heights.

\begin{figure}
  \includegraphics[width=0.6\textwidth]{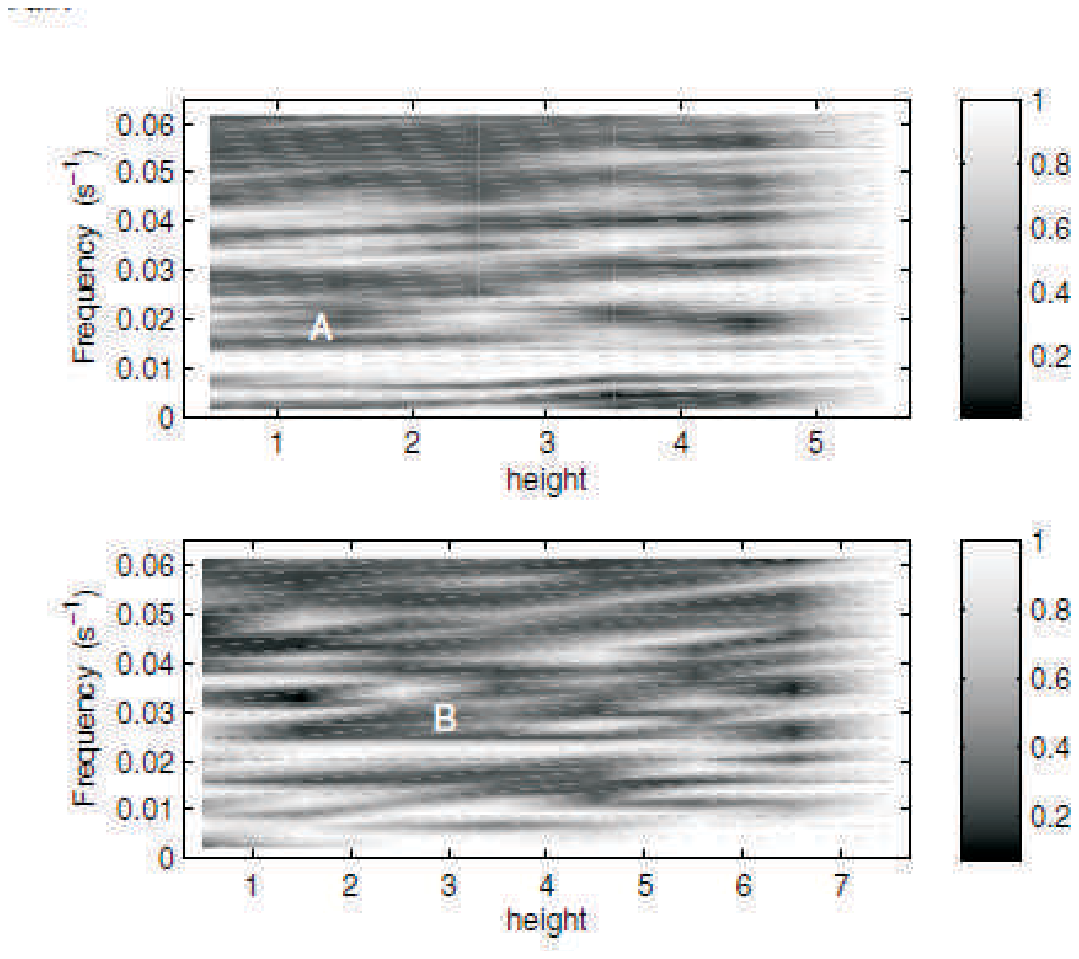}
  \includegraphics[width=0.6\textwidth]{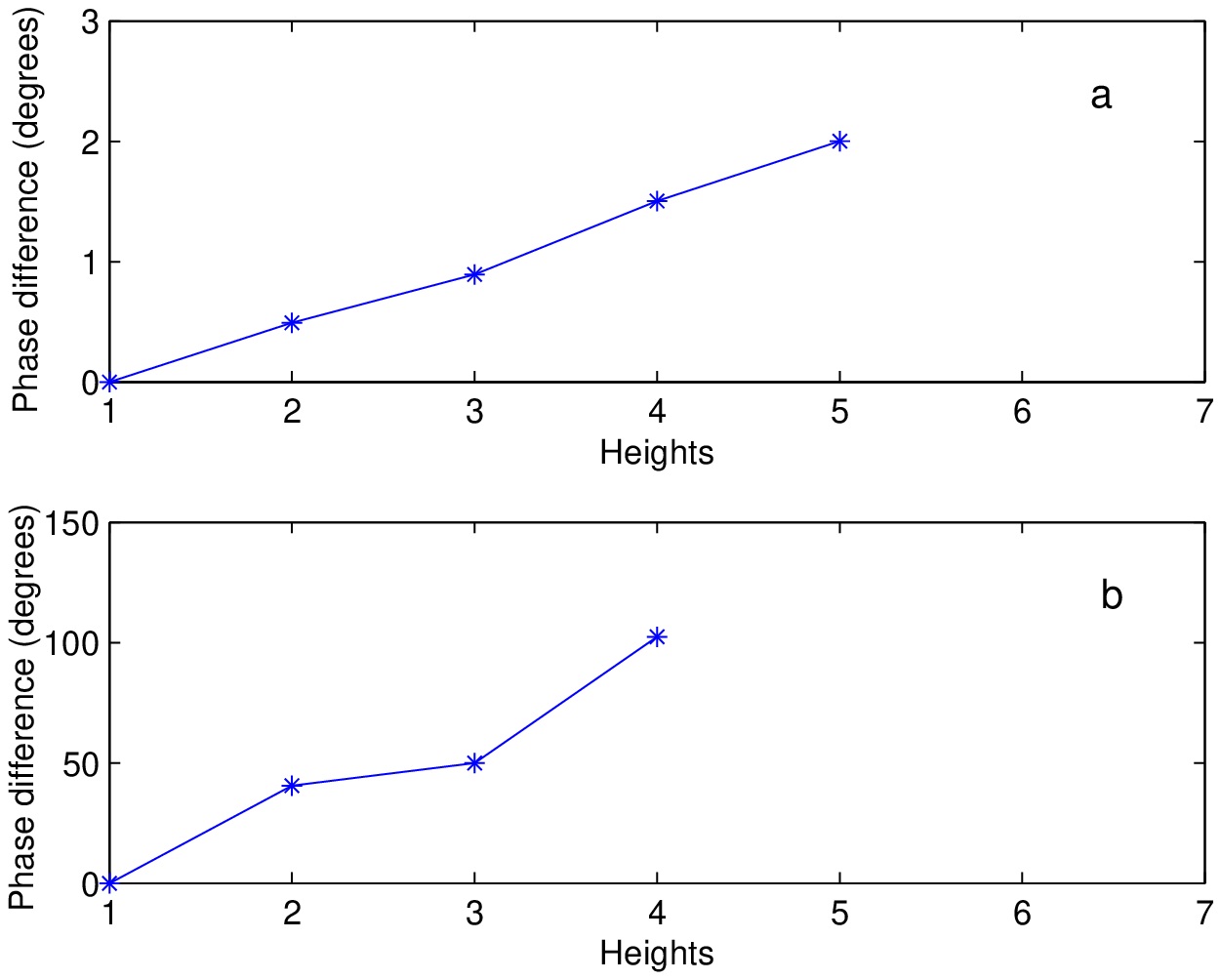}
\caption{Left: Fourier power expressed in confidence levels as
function of frequency and heights for two different spicules,
adapted from \cite{zaq07}. Brighter points correspond to higher
power, and darker points correspond to lower one. The label 1 on the
power scale (right plots) corresponds to the level of 100$\%$
confidence. Right: Relative Fourier phase as a function of height
for oscillations (a) with $\sim$80~s period in the first spicule
(feature A) and (b) 44 s period in the second spicule (feature B).
The distance between consecutive heights is $\sim$ 1 arc sec. }
\label{fig:16}       
\end{figure}

\cite{zaq07} presented a Fourier power as a function of frequency
and heights for two different spicules shown on Fig.~\ref{fig:16} (left
panel). There is clear evidence of persisting oscillations along the
full length of both spicules. The plot of the first spicule shows
the long white feature (feature A) located just above the frequency
0.01 s$^{-1}$. This is the oscillation with the period of $\sim$80~s
and it persists along the spicule. The most pronounced feature
(feature B) in the plot of the second spicule is colour-code indicated
by a long brighter trend located just above the frequency of 0.02 s$^{-1}$
and persisting along almost the whole spicule. This is the oscillation with a
period of 44~s.
Then, Zaqarashvili et al. calculated the relative Fourier phase between heights for
the most pronounced features (features A and B). Right panel of
Figure~\ref{fig:16} shows the relative Fourier phase as a function of heights
for (a) feature A and (b) feature B. In spite of the apparent linear behaviour of
the phase difference, there is practically almost no phase difference between
oscillations at different heights for feature A, which probably indicates a
standing-wave like pattern with period of $\sim$ 80-90 s. On the contrary, there is
a significant linear phase shift in plot (b), which indicates a propagating pattern
with an estimated period of $\sim$40-45 s. Therefore, the authors concluded that
the first spicule shows the standing-wave like pattern (or a wave propagation almost
along line of sight, which seems unlikely though cannot be ruled out) while the second
spicule shows a propagating wave pattern. Zaqarashvili et al. estimated the propagation
speed as $\sim$110 km s$^{-1}$.

Hence, based on these few observations we conclude that the
propagation speed of disturbances in solar limb spicules is quiet
high and may exceed 100 km s$^{-1}$. This indicates that the
disturbances are of magnetic origin and they propagate with
chromospheric Alfv\'en or kink speeds that exceeds the local sound
speed in spicules at these heights (but, note that the sound speed
outside spicules has almost coronal values).

\section{Possible interpretation of observed oscillations}
\label{sec:interpret}

It is clear that the observed transverse oscillations of spicule
axis can be explained and interpreted by the waves propagating along
the spicule. The two type of waves responsible for periodic
transverse displacement of spicule axis are: kink or Alfv\'en waves.
If spicules are modelled as plasma jets being shot along magnetic
flux tube, then the transverse oscillations could be caused by MHD
kink waves (\cite{kuk06,zaq07,erd07,aja09}). If a spicule is not
stable wave guide for the tube waves, then the oscillations can be
caused by Alfv\'en waves (\cite{dep07a}). Before we embark on the
interpretation of spicule oscillations, let us briefly summarise the
possible MHD modes and their main properties in an inhomogeneous
magnetised plasma under lower solar atmospheric conditions.

\subsection{MHD waves in a uniform magnetic cylinder in the lower solar atmosphere}
\label{ssec:mag_cyl_waves}

\begin{figure}[h!]
\begin{center}
\includegraphics[scale=0.4]{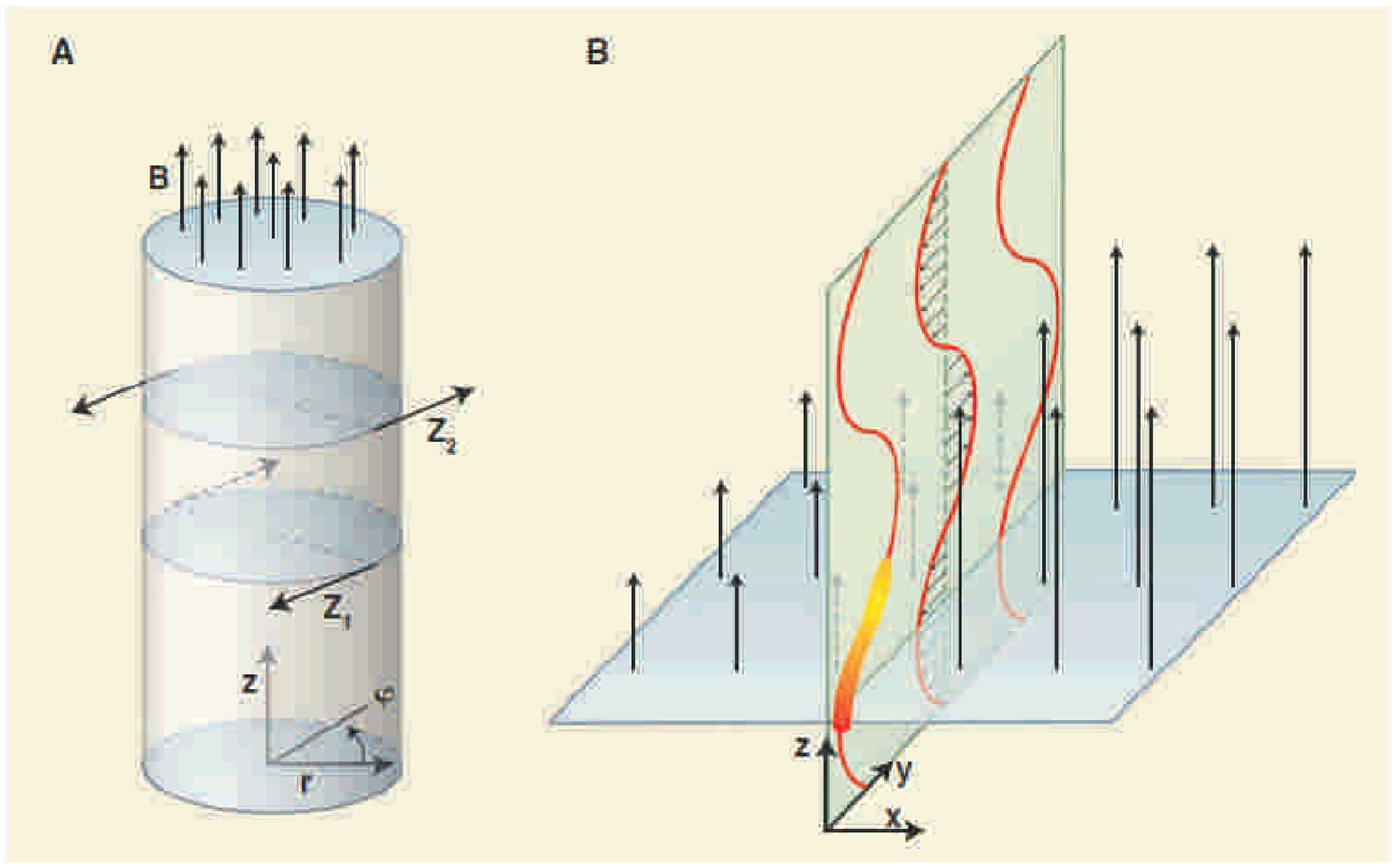}
\caption{\label{fig:tube_disc} {\it Left:} Magnetic flux tube
showing a snapshot of Alfv\'en wave perturbation propagating in the
longitudinal $z$-direction along field lines at the tube boundary.
At a given height the Alf\'enic perturbations are torsional
oscillations, i.e. oscillations are in the $\varphi$-direction,
perpendicular to the background field. Note that on the other hand
MHD kink waves would force the tube axis to oscillate. {\it Right:}
Snapshot showing Alfv\'en waves propagating along a magnetic
discontinuity. Again, the key feature to note is that Alfv\'enic
perturbations are {\it within} the magnetic surface ($yz$-plane) at
the discontinuity, perpendicular to the background field
($y$-direction), while the waves themselves propagate along the
field lines ($z$-direction). The MHD kink waves oscillate in
$xz$-plane in this geometry. Image adapted from \cite{erd07}. }
\end{center}
\end{figure}
To simplify the bewildering complexity of the dynamic solar
atmosphere, the concept of magnetic flux tubes is often used. In a
pioneering work by \cite{edw83}, using cylindrical coordinates, it
was derived the dispersion relations of MHD waves propagating in
cylindrical magnetic flux tubes. The main obstacle to be overcome
when introducing the concept of flux tubes is the conversion from
Cartesian to cylindrical coordinates. This change results involving
Bessel functions in the dispersion relation which are not yet
possible to be solved analytically without simplification, e.g.
through incompressibility or long and short wavelength
approximations. Let us summarise here the key steps of \cite{edw83}.
Consider a uniform magnetic cylinder of magnetic field
$B_0\mathbf{\hat{z}}$ confined to a region of radius $a$, surrounded
by a uniform magnetic field $B_e\mathbf{\hat{z}}$ (see
Figure~\ref{fig:tube_disc}a). To simplify the MHD equations we
assume zero gravity, there are no dissipative effects and all the
disturbances are linear and isentropic. Pressure (plasma and
magnetic) balance at the boundary implies that

\begin{equation}
p_0 + \frac{B_0^2}{2\mu_0}=p_e +\frac{B_e^2}{2\mu_0},
\label{eq:mcw.1}
\end{equation}
where $p_0$ and $p_e$ are the pressures inside and outside the tube.

\noindent
Linear perturbations about this equilibrium give the following pair of equations valid
inside the tube,

\begin{equation}
\frac{\partial^2}{\partial t^2}\Bigg( \frac{\partial^2}{\partial t^2}-(c_0^2+v_A^2)
\nabla^2\Bigg)\Delta+c_0^2v_A^2\frac{\partial^2}{\partial z^2}\nabla^2\Delta=0,
\label{eq:mcw.2}
\end{equation}

\begin{equation}
\Bigg(  \frac{\partial^2}{\partial t^2} -v_A^2\frac{\partial^2}{\partial z^2}\Bigg)\Gamma=0,
\label{eq:mcw.3}
\end{equation}

\noindent where $\nabla^2$ is the Laplacian operator in cylindrical
coordinates $(r,\phi,z)$ and

\begin{equation}
\Delta \equiv \mathrm{div}\mathbf{v},\quad \Gamma=\mathbf{\hat{z}}\cdot\mathrm{curl}\mathbf{v}
\label{eq:mcw.4}
\end{equation}

\noindent for velocity $\mathbf{v}=(v_r,v_\phi, v_z)$. A similar
pair of equations to (\ref{eq:mcw.2}) and (\ref{eq:mcw.3}) are valid
outside the tube. Fourier analysing we let

\begin{equation}
\Delta=R(r)exp[i(\omega t + n\phi + kz). \label{eq:mcw.5}
\end{equation}

\noindent
Then equations (\ref{eq:mcw.2}) and (\ref{eq:mcw.3}) give Bessel's  equation satisfied by
$R(r)$ as follows

\begin{equation}
\frac{\mathrm{d}^2R}{\mathrm{d}r^2} +\frac{1}{r}\frac{\mathrm{d} R}{\mathrm{d}r}-
\Bigg( m_0^2 + \frac{n^2}{r^2}\Bigg)R=0,
\label{eq:mcw.6}
\end{equation}

\noindent
where

\begin{equation}
m_0^2=\frac{(k^2c_0^2-\omega^2)(k^2v_A^2-\omega^2)}{(c_0^2+v_{A}^2)(k^2c_{T}^2-\omega^2)}.
\label{eq:mcw.7}
\end{equation}

\noindent We have used the notation $v_{A}$ for the Alfv\'en speed,
$c_0$ for the sound speed and $c_T$ for the characteristic tube
speed (sub-Alfv\'{e}nic), where $c_T=c_0v_A/(c_0^2+v_A^2)^{-1/2}$.
To obtain a solution to (\ref{eq:mcw.6}) bounded at the axis $(r=0)$
we must take

\begin{equation}
R(r)=A_0\left\{\begin{array}{ll}
I_n(m_0r), & m_0^2>0\\
J_n(n_0r), & n_0^2=-m_0^2>0
\end{array}\right\} (r<a),
\label{eq:mcw.8}
\end{equation}

\noindent where $A_0$ is an arbitrary constant and $I_n$, $J_n$ are Bessel
functions, see e.g. \cite{abr67}, of order $n$. For a mode locked to the waveguide it is
required that no energy propagates to or from the cylinder in the external region, i.e.
the waves are evanescent outside the flux tube. Therefore we take

\begin{equation}
R(r)=A_1K_n(m_er), \quad r>a,
\label{eq:mcw.9}
\end{equation}

\noindent where $A_1$ is a constant and

\begin{equation}
m_e^2=\frac{(k^2c_e^2-\omega^2)(k^2v_{Ae}^2-\omega^2)}{(c_e^2+v_{Ae}^2)(k^2c_{Te}^2-\omega^2)},
\label{eq:mcw.10}
\end{equation}

\noindent which is taken to be positive (no leaky waves). Here,
$c_e$ stands for the sound speed outside the tube. Since we must
have continuity of velocity component $v_r$ and total pressure at
the cylinder boundary $r=a$, this yields the dispersion relations

\begin{equation}
\rho_0(k^2v_A^2-\omega^2)m_e\frac{K_n'(m_ea)}{K_n(m_ea)}=
\rho_e(k^2v_{Ae}^2-\omega^2)m_0\frac{I_n'(m_0a)}{I_n(m_0a)},
\label{eq:mcw.11}
\end{equation}

\noindent for surface waves $(m_0^2>0)$ and

\begin{equation}
\rho_0(k^2v_A^2-\omega^2)m_e\frac{K_n'(m_ea)}{K_n(m_ea)}=
\rho_e(k^2v_{Ae}^2-\omega^2)n_0\frac{J_n'(n_0a)}{J_n(n_0a)},
\label{eq:mcw.12}
\end{equation}

\noindent for body waves $(m_0^2=-n_0^2<0)$. The axisymmetric
sausage mode is given by $n=0$, while the well-observed kink mode
(non-axisymmetric) is given by $n=1$. Modes with $n>1$ are called
flute modes. Although the dispersion relations (\ref{eq:mcw.11}) and
(\ref{eq:mcw.12}) are complicated, finding the phase speed for e.g.
kink waves with photospheric parameters simplifies matters
considerably.

\begin{figure}[h!]
\begin{center}
\includegraphics[width=0.7\textwidth]{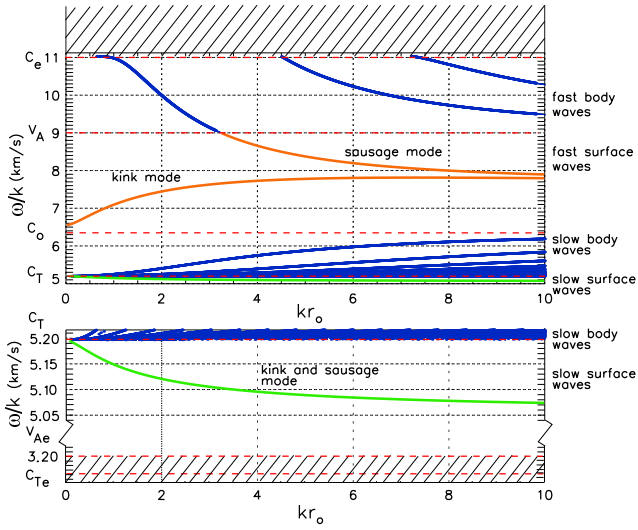}
\caption{\label{fig:dr_photo1}The solution of the dispersion relations
(\ref{eq:mcw.11})-(\ref{eq:mcw.12}) in terms of phase speed ($\omega/k$) of modes
under photospheric conditions $c_{e}>v_{A}>c_{0}$ (all speeds are in km/s).
The slow band is zoomed (lower panel). Image adapted from \cite{erd08}.}
\end{center}
\end{figure}

\begin{figure}[h!]
\begin{center}
\includegraphics[width=0.7\textwidth]{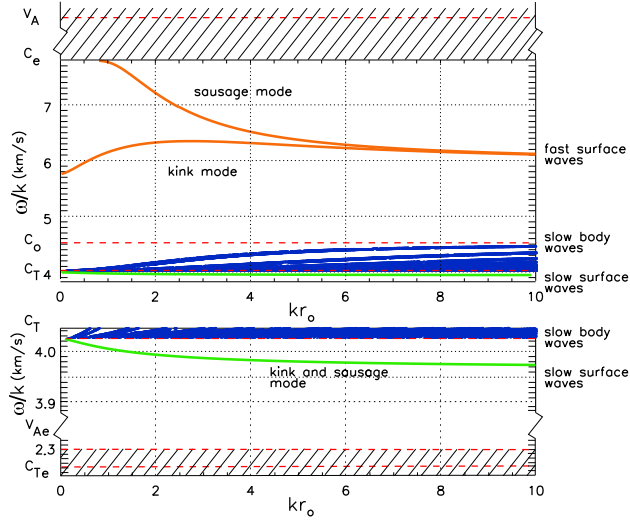}
\caption{\label{fig:dr_photo2}Similar to Fig.~\ref{fig:dr_photo1} but for
$v_{A}>c_{e}>c_{0}$. Image adapted from \cite{erd08}.}
\end{center}
\end{figure}

Under lower solar atmospheric conditions, characterised by $c_{e}>v_{A}>c_{0}$, possibly
representative for spicules, sunspots or pores both the slow and fast bands have surface
and body modes, respectively. The slow MHD waves are in a narrow band since $c_0\approx c_T$.
The slow body waves are almost non-dispersive, whereas the almost identical slow surface
sausage and kink modes are weakly dispersive (bottom zoomed out panel in Fig.~\ref{fig:dr_photo1}).
On the other hand, if the characteristic speeds of a lower solar
atmospheric flux tube render as $v_{A}>c_{e}>c_{0}$, the fast body
modes do not exist since they would become leaky mode solutions of
the dispersion relations (\ref{eq:mcw.11})-(\ref{eq:mcw.12}). The
MHD modes propagating along the flux tube in this case are shown in
Fig.~\ref{fig:dr_photo2}. Note, there is little evidence, whether
flux tubes modelling spicules, are characterised by
$c_{e}>v_{A}>c_{0}$ or $v_{A}>c_{e}>c_{0}$.

In the incompressible limit, suitable for the description of kink
waves and oscillations in their linear limit, $(c_0^2\rightarrow
\infty, c_e^2\rightarrow \infty), m_0$ and $m_e$ become $|k|$. The
kink and sausage modes are then given explicitly after some algebra.
It is noted that the phase speed for the kink mode is not monotonic
as a function of $k$ but has a maximum/minimum and the sausage mode
is monotonically increasing/decreasing. This max/min feature of the
kink mode is absent in the slab case, so it can be deduced to be a
reflection of the geometry of the magnetic field. On the other hand,
the incompressible Alfv\'en waves in magnetic tubes are polarized in
the $\phi$ direction and do not lead to the displacement of the tube
axes (see Figure~\ref{fig:tube_disc}a). The propagation of torsional
Alfv\'en waves in vertical magnetic tubes have been studied by
\cite{hol81} and \cite{hol82}.

Particularly interesting processes may take place near the heights,
where the Alfv\'en and sound waves have similar values of phase
speed. This area with $\beta=1$ ($\beta=8\pi p_0/B^2_0$), may locate
somewhere between the photosphere and chromosphere.
\cite{bog02,bog03} have performed 2D numerical simulations in
isothermal atmosphere and have shown the coupling between different
MHD wave modes near this area. \cite{zaq06} also show that the
nonlinear coupling between Alfv\'en and sound waves may take place
there.

Last but not least, for completeness and for the benefit of
interested readers, we note that kink MHD wave propagation under
solar coronal conditions is discussed in details by \cite{rud09} in
this Volume.

\subsection{MHD kink waves}
\label{ssec:kink_waves}

\begin{figure}[h!]
     \includegraphics[width=0.5\textwidth]{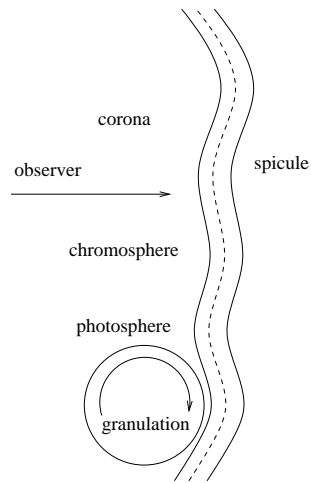}
\caption{Schematic picture of propagating kink waves in spicules, adapted
from \cite{kuk06}.}
\label{fig:17}       
\end{figure}

Equipped with a clear understanding of the differences between kink
and Alfv\'enic perturbations, as described above, let us now return
to plausible interpretations of periodic spicular motions. Spicules
are much denser than the surrounding medium above the height of 2000
km (see section 2.2), therefore they may be considered as magnetic
tubes and the MHD wave theory can be applied in {\it some} format as
outlined in the Sec.~\ref{ssec:mag_cyl_waves}. Then, the observed
periodic transverse displacement of the axis probably is due to the
propagation of kink waves (see Figs.~\ref{fig:tube_disc} and
\ref{fig:17}). Transverse kink waves can be generated in
photospheric magnetic tubes by buffeting of granular motions
(\cite{rob79,spr81,osi99}). The waves may then propagate through the
stratified chromosphere (see, e.g. \cite{har08,har09}) and lead to
the observed oscillations (\cite{kuk06,zaq07}). The recent
theoretical paper by \cite{aja09} also shows that the kink waves
with periods of $\sim$ 80-120 s may propagate in spicules at higher
heights. Indeed, the oscillations of Doppler velocity in spicules
with similar periods were reported by \cite{zaq07}. If the velocity
of the kink wave is polarized in the plane of observation, then it
results in the Doppler shift of the observed spectral line
(\cite{nik67,pas68,kul78,gad82,kul83,has84,pap94,xia05,kuk06,zaq07}).
However, if the velocity is polarized in the perpendicular plane
then it results the visible displacement of spicule axis along the
limb (\cite{nik67,nik71,gad82,pap94}).
Kink wave propagation along a vertical thin magnetic flux tube
embedded in the {\it stratified} field-free atmosphere is governed by the
Klein-Gordon equation (\cite{rae82,spr83,has99,zaq08,har08,har09}; the latter authors
have even considered a dissipative medium resulting in a governing equation of the type of
Klein-Gordon-Burgers equation)
\begin{eqnarray}
{{\partial^2 Q}\over {\partial z^2}}-{{1}\over c^2_k}{{\partial^2
Q}\over {\partial t^2}} - {{\Omega^2_k}\over c^2_k} Q =0,
\label{eq:kw.1}
\end{eqnarray}
where $Q=\xi(z,t)\exp{(-z/4\Lambda)}$,
$c_k=B_0/\sqrt{4\pi(\rho_0+\rho_e)}$ is the kink speed, $\Lambda$ is
the density scale height and $\Omega_k=c_k/4\Lambda$ is the
gravitational cut-off frequency for isothermal atmosphere
(temperature inside and outside the tube is assumed to be the same
and homogeneous). Here $\xi(z,t)$ is the transversal displacement of
the tube, $B_0(z)$ is the tube magnetic field, $\rho_0(z)$ and
$\rho_e(z)$ are the plasma densities inside and outside the tube
respectively (the magnetic field and densities are functions of $z$,
while the kink speed $c_k$ is constant in the isothermal
atmosphere).

Eq.~(\ref{eq:kw.1}) yields simple harmonic solutions $\exp[{i(\omega t \pm k_z
z)}]$ with the dispersion relation
\begin{eqnarray}
\omega^2 - \Omega^2_k= c^2_k k^2_z,
\label{eq:kw.2}
\end{eqnarray}
where $\omega$ is the wave frequency and $k_z$ is the wave number.
The dispersion relation shows that waves with higher frequency
than $\Omega_k$ may propagate in the tube, while the lower frequency
waves are evanescent.

\begin{figure}[ht!]
     \includegraphics[width=0.5\textwidth,angle=270]{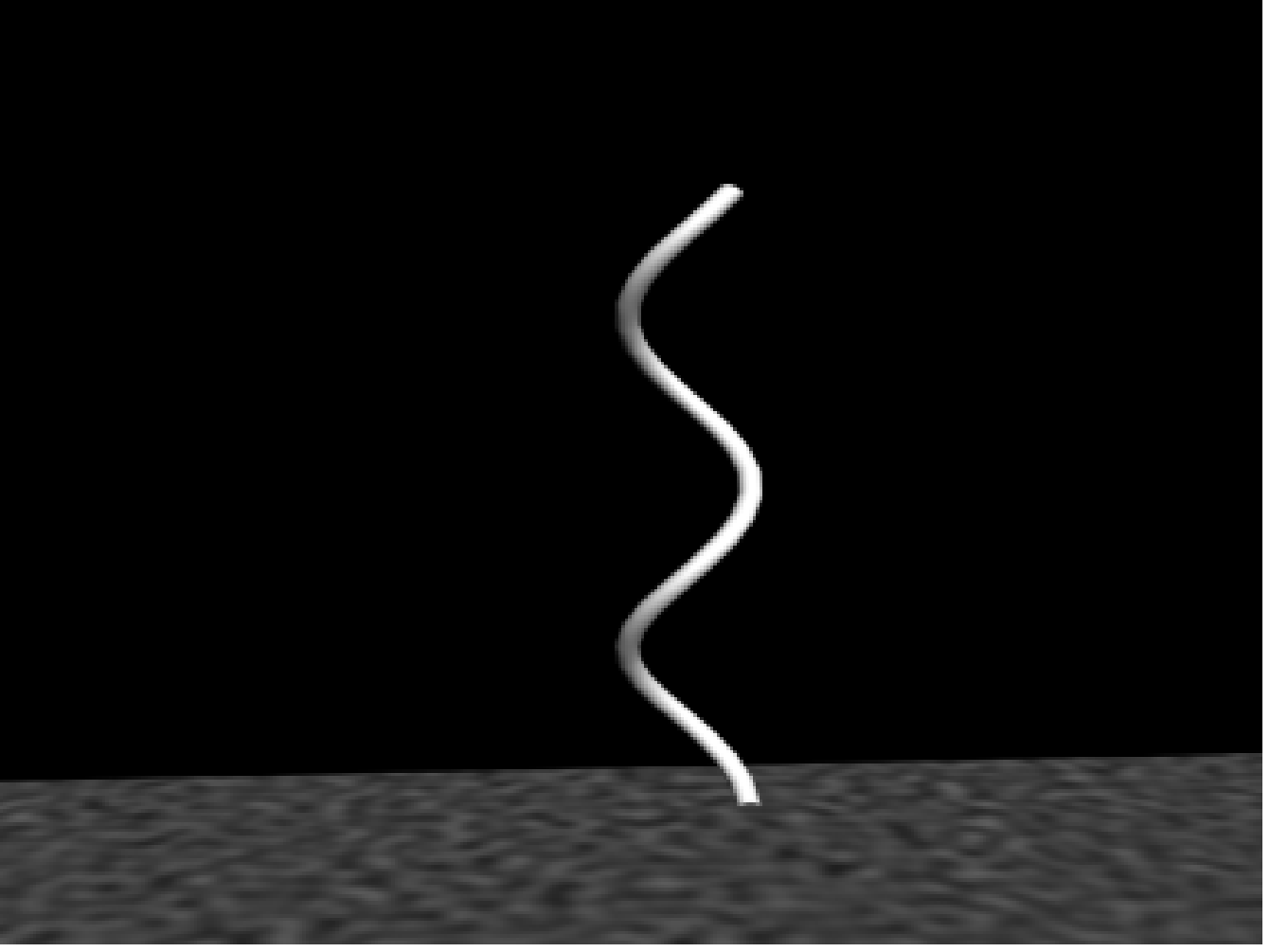}
\caption{Helical kink wave in a thin magnetic flux tube, adapted from
\cite{zaq08}.}
\label{fig:18}       
\end{figure}

Kink waves cause the transverse displacement of the entire tube. The
displacement of tube in a simple harmonic kink wave is polarized
arbitrarily and the polarization plane depends on the excitation
source. The superposition of two or more kink waves polarized in
different planes may give rise to the complex motion of the tube.
The process is similar to the superposition of two plane
electromagnetic waves, where the waves with the same amplitudes lead
to the circular polarization, while the waves with different
amplitudes lead to the elliptical polarization. Consider, for
example, two harmonic kink waves with the same frequency but
polarized in the $xz$ and $yz$ planes: $A_x=A_{x0}\cos(\omega t +
k_z z)$ and $A_y=A_{y0}\sin(\omega t + k_z z)$. The superposition of
these waves sets up {\it helical waves} with a circular polarization
if $A_{x0}=A_{y0}$ (\cite{zaq08}). As a result, the tube axis
rotates around the vertical, while the displacement remains constant
(Fig.~\ref{fig:18}). If $A_{x0} \not= A_{y0}$ then the resulting
wave is elliptically polarized. The superposition of few harmonics
with different frequencies and polarizations may lead to an even
more complex motion of the tube axis.

\subsection{Alfv\'en waves}
\label{ssec:alf_waves}

\begin{figure}[h!]
     \includegraphics[width=0.9\textwidth]{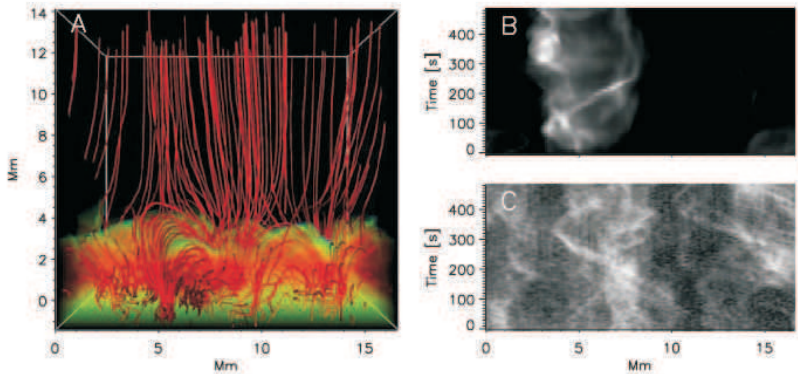}
\caption{Comparison between observations and simulations of Alfv\'en
waves from \cite{dep07a}. A: a snapshot from 3D radiative MHD
simulation. B: space-time plot from simulations. C: space-time cut
from observations. }
\label{fig:19}       
\end{figure}

\cite{dep07a} suggested that the transverse displacement of spicule
axis can be explained by the propagation of Alfv\'en waves excited
in the photosphere by granular motions or global oscillation
patterns. They performed self-consistent 3D radiative MHD
simulations ranging in the vertical direction from the convection
zone up to the corona. A snapshot from the simulations is presented
on the Fig.~\ref{fig:19} (panel A). Their analysis shows that the
field lines (red lines) in the corona, transition region, and
chromosphere are continuously shaken and carry Alfv\'en waves.
Panels B and C are time-distance plots from the simulations and
observations, respectively. From the comparison between simulations
and observations the authors concluded that the period of Alfv\'en
waves should lay between 100 and 500 s. However, they suggested that
very long-lived macro spicules show some evidence of Alfv\'en waves
with longer periods between 300 and 600 s. \cite{dep07a} also claim
that their observations and simulations do not show the spicules as
stable wave guide for kink waves. Therefore they argue that
volume-filling Alfv\'en waves cause the swaying of magnetic field
lines back and forth leading to the visible displacement of spicule
axis. These claims are debated by \cite{erd07}: "However, these
observations also raise concerns about the applicability of the
classical concept of a magnetic flux tube in the apparently very
dynamic solar atmosphere, where these sliding jets were captured. In
a classical magnetic flux tube, propagating Alfv\'en waves along the
tube would cause torsional oscillations (see
Fig.~\ref{fig:tube_disc}a in this paper earlier). In this scenario,
the only observational signature of Alfv\'en waves would be spectral
line broadening. Hinode/SOT does not have the appropriate
instrumentation to carry out line width measurements\footnote{We
need to clarify here that, of course, Hinode/SOT {\it can} measure
line width, however, not with the desired resolution and in the
desired wavelength in the context of spicule oscillations, as
opposed to e.g. the Rapid Oscillations in the Solar Atmosphere
(ROSA) instrument mounted on the Dunn Solar Telescope, NSO,
Sacramento Peak.}. On the other hand, if these classical flux tubes
did indeed exist, then the observations of \cite{dep07a} would be
interpreted as kink waves (i.e., waves that displace the axis of
symmetry of the flux tube like an S-shape). More detailed
observations are needed, perhaps jointly with
STEREO\footnote{Although STEREO has a limited spatial capacity for
spicule observations, it can give a global view of the related MHD
waveguide.}, so that a full three-dimensional picture of wave
propagation would emerge."

\subsection{Kink vs Alfv\'en waves}
\label{ssec:kink_vs_alf}

It is important to determine accurately which type of waves are
responsible for the transverse displacement of the tube axis
(\cite{erd07,jes09}). The existence and main features of wave modes
depend on the properties of the medium where the waves propagate in
(see Sec.~\ref{ssec:mag_cyl_waves}). If spicules represents a
magnetic flux tube, then MHD wave theory may permit the propagation
of kink and torsional Alfv\'en waves. The kink waves are global tube
waves and cause the displacement of the tube as a whole. However,
most importantly let us emphasise once more that torsional Alfv\'en
waves do not lead to the displacement of tube axis
(\cite{erd07,van08,jes09}). Therefore, torsional Alfv\'en waves are
unlikely to be the reason for the observed oscillations. However,
{\it global} Alfv\'en waves, which may indeed fill a significant
volume in and around spicules, may cause the global transverse
oscillations of magnetic field lines. In this latter scenario
spicules may just simply follow the oscillations of field lines and
this is what \cite{dep07a} suggest. However, it must be mentioned
here, that spicules may respond to the oscillation of magnetic field
lines slower than the surrounding plasma due to their higher density
i.e. larger inertia. This point is not taken
into account by \cite{dep07a} and it needs to be addressed in future models.  \\
However, there are two possible difficulties associated with the
Alfv\'en wave scenario. Firstly, if the oscillations of spicules are
due to global Alfv\'en perturbations, then the neighbouring spicules
should show a coherent oscillation. However, Hinode movies indicate
the opposite, i.e. spicule movements are random and there is no sign
of coherent oscillations. This incoherence partly can be caused due
to the fact that spicules seen at the limb are located in different
parts along the line of sight. However, detailed analysis still can
uncover the coherent oscillations and this point needs to be
addressed in future. Secondly, it is not yet clear how the
volume-filling Alfv\'en waves would be generated in the photosphere,
where the magnetic field is rooted and concentrated in flux tubes.
In this regards, it seems to be more plausible that the transverse
perturbations propagate upwards in the form of kink waves, which may
be transformed into Alfv\'en-like waves in the chromosphere, where
the magnetic field rapidly expands. The Alfv\'en waves also can be
generated near $\beta=1$ region due to various wave
coupling processes (\cite{bog02,bog03,zaq06}). \\
The kink wave scenario also has its own difficulties as the
photospheric flux tubes may be expanded in the chromosphere as we
already noted above. This expansion may cause certain difficulties
for the kink wave to propagate. However, plasma $\beta$ becomes less
than unity in higher heights, which means that the concept of
magnetic tube is changing compared to the photospheric conditions.
Now, the magnetic tube means the higher concentration of density.
This is exactly the case for spicules, as their density can be up to
two order magnitude higher, than in the surrounding coronal plasma.
Therefore, the spicules can be wave guide of the kink waves (at
least, in classical spicules) although \cite{dep07a} claim opposite.
Propagation of transverse pulse in the wave guide of enhanced
density was recently studied by \cite{doo08}. They show that the
slab with enhanced density essentially trap the initial transverse
pulse. It must be mentioned that the rapid disappearance of type II
spicules in Ca II of Hinode/SOT does not immediately mean that
spicules are not stable wave guides. The disappearance can be caused
due to the increase of temperature inside spicule. But the tube
(i.e. higher concentration of density) still may remain. In this
regards, \cite{erd07}, in the context of prominence oscillations,
showed that kink waves can be easily guided.  For the recently
developed theory of transversal waves and oscillations in
gravitationally and magnetically stratified flux tubes see, e.g.
\cite{ver08,rud08,and09}. The problem whether the spicule
displacements are Alfv\'enic or kink waves is currently under debate
and more sophisticated observations complemented by numerical
investigations are needed for a satisfactory solution
(\cite{erd07}).

\subsection{Transverse pulse}
\label{ssec:trans_pulse}

It must be recognised that simple harmonic waves can hardly be excited in the dynamic
solar photosphere. A more realistic process of wave excitation is the impulsive buffeting
of granules on an anchored magnetic flux tube, which may easily generate transverse pulses.
Such pulses may propagate upwards in the stratified atmosphere and leave the "wake" oscillating
at cut -off frequency of kink waves (\cite{zaq08}). The wake may be also responsible for the
observed transverse oscillations of spicule axes.

For the sake of simplicity, let us consider the simplest impulsive
forcing in both time and coordinate. Then Eq.~(\ref{eq:kw.1}) looks as
(\cite{zaq08})
\begin{equation}
{{\partial^2 Q}\over {\partial z^2}}-{{1}\over c^2_k}{{\partial^2
Q}\over {\partial t^2}} - {{\Omega^2_k}\over c^2_k} Q =-
A_0\delta(t)\delta(z),
\label{eq:tp.1}
\end{equation}
where $z>-\infty$, $t>0$, $A_0$ is a constant and the pulse is set off at
$t=0$, $z=0$.

\begin{figure}
     \includegraphics[width=0.8\textwidth,angle=0]{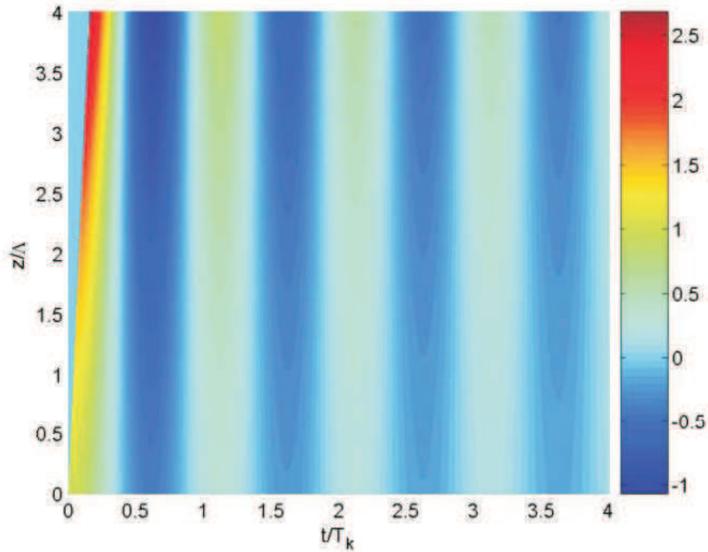}
\caption{Propagation of transverse kink pulse in stratified vertical
magnetic flux tubes, adapted from \cite{zaq08}. A rapid propagation of the
pulse is seen, which is followed by the oscillating wake. The pulse
propagates with the kink speed and the wake oscillates with the
cut-off period. The amplitude of the pulse (and wake) increases with
height due to the decreasing density.}
\label{fig:20}       
\end{figure}

The solution of this equation can be written, after e.g. (\cite{mor53}),
\begin{equation}
Q=A_0c_k\delta\left(t-{z\over c_k} \right ) - {{c_kA_0}\over
{2}}J_0\left [\Omega_k\sqrt{t^2 - {{z^2}\over {c^2_k}}} \right
]H\left [\Omega_k \left ( t- {{z}\over {c_k}} \right )\right ],
\label{eq:tp.2}
\end{equation}
where $J_0$ and $H$ are Bessel and Heaviside functions, respectively.
Eq.~(\ref{eq:tp.2}) shows that the wave front propagates with the kink speed
$c_k$ (the first term), while the wake oscillating at the cut-off
frequency $\Omega_k$ is formed behind the wave front (the second
term) and it decays as time progresses \citep{rae82,spr83,has99,har08,har09}.
Note, the actual mathematical form of the governing equation and its solution for
kink and longitudinal oscillations in a gravitationally stratified and anchored
magnetic flux tube is very similar, see e.g. \cite{sut98,mus01,bal06} for more
details.\\
Fig.~\ref{fig:20} shows the plot of the transverse displacement
$\xi(z,t)=Q(z,t)\exp{(z/4\Lambda)}$, where $Q$ is expressed by the
second term of Eq.~(\ref{eq:tp.2}). A rapid propagation of the pulse is found,
which is followed by the oscillating wake (the time is normalized by
the cut-off period $T_k=2\pi/\Omega_k$). Just after the propagation
of the pulse, the tube begins to oscillate with the cut-off period
at each height. The amplitudes of pulse and wake increase upwards
due to the density reduction, but the oscillations at each height
decay in time. A very similar and resembling behaviour was found by
\cite{fle89,fle91,sch98,erdetal07,mal07} for longitudinal oscillations.

Hence, the transverse and impulsive action on the magnetic tube at
$t=0$ near the base of the photosphere (as set at $z=0$) excites the
upward propagating kink pulse, while the tube in the photosphere
oscillates at the photospheric kink cut-off period, which can be
estimated as $\sim$8 min, using the photospheric scale height of 125
km. Hence, the magnetic tube will oscillate with $\sim$8 min period
in the photosphere. When the pulse penetrates into the higher
chromosphere, where the cut-off period is changed, this will
influence the propagation of the pulse. Spicules with higher density
concentrations than the ambient plasma may guide kink waves with the
phase speed of 25 km/s, which yields the cut-off period of about 500
s. Note, the Alfv\'en cut-off period can be as short as 250 s.
Therefore, the transverse pulse may set up the oscillating wake in
the chromosphere with the period of 250-500 s.

\begin{figure}[h!]
     \includegraphics[width=0.7\textwidth,angle=0]{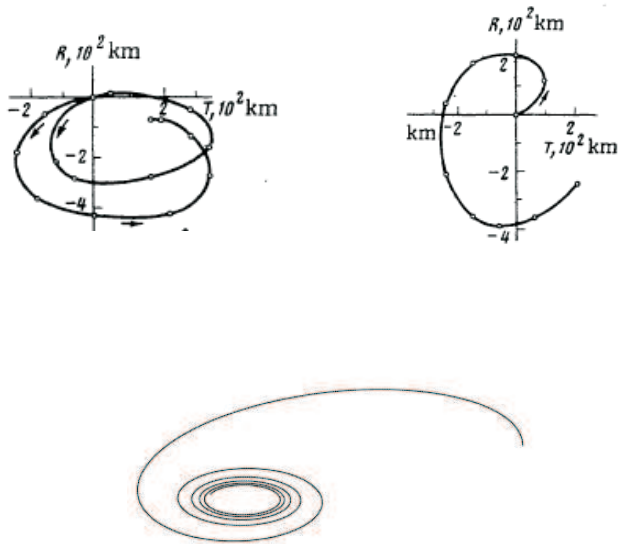}
\caption{Upper panels: Observed trajectories of the motion of two
distinct spicules, adapted from \cite{gad82}. Lower panel: Simulated
helical motion of the tube axis at the height of 250 km measured from the
solar surface, due to the propagation of two consecutive transverse pulses
polarized in perpendicular planes, adapted from \cite{zaq08}. }
\label{fig:21}       
\end{figure}

It is most likely, that the anchored magnetic flux tube undergoes
granular buffeting from many directions. Therefore, the
superposition of consecutive pulses may set up a helical motion of
tube axis with the cut-off period (\cite{zaq08}). The helical motion
of spicule axis first has been observed by \cite{gad82} from
simultaneous observation of Doppler velocity and visual
displacements. The recent Hinode/SOT movies also show complex
motions of spicule axis. Fig.~\ref{fig:21} (upper panels) show the
observed trajectories of spicule axis with respect to the
photosphere, adopted from \cite{gad82}. The lower panel shows the
superposition of two wakes at the height of 250 km above the
photosphere, numerically simulated by \cite{zaq08}, by solving the
Klein-Gordon governing equation. The first wake corresponds to the
pulse imposed along the $x$-direction and the second wake is the
result of another pulse generated in the $y$-direction with a
different amplitude. The observed period of helical motions is in
the range of 3-6 mins, which may correspond well with the kink
cut-off period in the chromosphere. A very similar phenomenon was
recently observed in fibril orientation by \cite{koz07}.
Therefore, the oscillation of wakes behind a transverse pulse may
explain the visible transverse displacement of spicule axis observed
by \cite{dep07a} with Hinode/SOT. The cut-off period is similar or
slightly shorter than the mean life time of spicules, therefore the
oscillations are difficult to detect as it is noted by
\cite{dep07a}. It must be mentioned, however, that the
photosphere/chromosphere is in much more complex and dynamic state
than it is described by this simple theoretical approach. Therefore,
it is desirable to perform more sophisticated numerical simulations
of transverse pulse propagation from the photosphere up to the
corona.

\section{Summary of main results and targets for future work}
\label{sec:summ_future}

Spicules are one of the most plausible tracers and trackers of the
energy coupling and energy transport from the lower solar
photosphere towards the upper corona by means of MHD waves. These
waves may induce the oscillatory phenomena in the chromosphere,
which are frequently detected in limb spicules. Periodic
perturbations, e.g. in forms of oscillations, are observed by both
spectroscopic and imaging observations. Let us summarize the main
observed oscillatory phenomena (see also Table~\ref{tab:2}):
\begin{itemize}

\item Oscillations in limb spicules are more frequently observed in
Doppler shifts and in the visible displacement of spicule axis, which probably
indicate the presence of transversal motions of spicules as a whole.

\item The observed oscillation periods can be formally divided into two groups: those with shorter
($<$2-min) and those with longer ($\geq$2-min) periods.

\item The most frequently observed oscillations are with period ranges of $50-110$
s and $3-7$ min.

\item The propagation of the actual oscillations is rather difficult to detect. However,
the relative Fourier phase between oscillations at different heights indicates the
propagation speed of $\sim$ 110 km s$^{-1}$. In some oscillations, perhaps caused by standing
patterns, waves seem to be present with very high phase speeds ($>$ 300 km s$^{-1}$).

\end{itemize}

The observed oscillations in spicules are most likely due to the
propagation of transverse waves from the photosphere towards the
corona. There are several possible interpretations of these
oscillatory phenomena:

\begin{itemize}

\item Kink waves propagating along slender magnetic flux tubes, where spicules are
formed on field lines close to the axis. The kink waves lead to the transverse oscillations
of spicules as a whole.

\item Volume-filling Alfv\'en waves propagating in surroundings of spicules. The Alfv\'en waves result in the
oscillation of the ubiquitous magnetic field lines. These oscillations force the
spicules to periodic displacement of their axes.

\item Transverse pulses excited in the photospheric magnetic flux tube by means of
buffeting of granules. The pulses may propagate upwards in the
stratified atmosphere and leave "wakes" behind, which oscillate at
the kink cut-off frequency of the stratified vertical magnetic flux
tube. The wakes can be responsible for the observed $\geq$~3-min
period transversal oscillations in limb spicules.

\end{itemize}

It is important to estimate the energy flux stored in spicule
oscillations. If the oscillations are caused by the waves excited by
granulation, then the energy flux of waves should be in the range of
the energy flux of granulation. The energy flux stored in an initial
transverse pulse at the photosphere is $F \sim n_e c_k v^2_g$, where
$v_g$ is the granular velocity, say 1-2 km s$^{-1}$. Then, for the
photospheric electron density of 2 ${\cdot}10^{17}$ cm$^{-3}$ and
Alfv\'en speed of $10$ km s$^{-1}$, the estimated energy flux is
$\sim 5{\cdot}10^9$ erg
cm$^{-2}$ s$^{-1}$ (taking $v_g$ as 1.25 km/s).\\
The same energy flux under spicule conditions (i.e. electron density
of 2 ${\cdot}10^{11}$ cm$^{-3}$ and Alfv\'en speed of 100 km
s$^{-1}$) requires the wave velocity in spicules to be about 385
km/s. The expansion of magnetic tube with height may alter the
estimation. However, spicule density is almost two magnitude higher
compared to the other part of the tube. Therefore, even if spicules
occupy only $1\%$ of the tube cross section, the energy storied in
spicule oscillations is comparable to the oscillation energy of
remaining part of the tube. Therefore, even if the wave transmission
efficiency is less than 100 $\%$, the observed velocity is much
lower than expected. This discrepancy can be resolved if the
oscillations are caused by the wake behind a transversal pulse: in
this case {\it almost the entire energy of the initial perturbation
is carried by the pulse, while the energy of the wake is much
smaller}. Therefore, even if the filling factor of magnetic tubes is
$1\%$, the energy flux carried by pulses is more than enough to heat
the solar chromosphere/corona.

\subsection{Targets for future observations}

More observations from space satellites and ground based coronagraphs are needed for a better
and conclusive understanding of oscillatory events in solar limb spicules. There are few
highlighted targets for future observations:

\subsubsection{Phase relations between oscillations in neighbouring spicules}

It is important to perform an analysis of phase relations between
oscillations in neighbouring spicules. If spicule oscillations are
caused by global Alfv\'en waves or they are related to global
photospheric oscillations, then the transverse displacement of
spicules should show some spatial coherence i.e. a few neighbouring
spicules should move in phase along the limb. Superposition of
different spicule groups located along the line of sight could
complicate the task, but careful analysis still may reveal some
coherence. Hinode/SOT time series in Ca II H and H$\alpha$ lines
seem to be an excellent data for such analysis work.

\subsubsection{Phase relations between oscillations at different
heights}

A study the phase relation of transverse displacements of a
particular spicule at different heights would allow us to infer whether
the oscillations are due to standing or propagating wave patterns.
Phase delays between different heights would determine the phase
speed of perturbations, thus, the physical nature of the waves. Spectroscopic
consecutive height series from ground based coronagraphs or time series of images
from e.g. Hinode/SOT would allow to infer the wave length, phase speed or frequency
of oscillations. These latter important diagnostic parameters may be then used to
develop {\it spicule seismology} as suggested by \cite{zaq07}.

\subsubsection{Propagation of transverse pulse}

Propagation of transverse pulses can be traced through a careful
analysis of time series from e.g. Hinode/SOT. It is also important
to search for oscillations of spicule axis at the kink or Alfv\'en
cut-off frequency as estimated in the chromosphere. Polar macro-spicules
are probably the best targets for such work, as their life time is long
enough when compared to the cut-off period.

%



%

\begin{acknowledgements}
This paper review was born out of the discussions that took place at
the International Programme "Waves in the Solar Corona" of the
International Space Science Institute (ISSI), Bern. The authors
thank for the financial support and great hospitality received
during their stay at ISSI. TZ acknowledges Austrian Fond zur
F\"{o}rderung der wissenschaftlichen Forschung (project P21197-N16)
and Georgian National Science Foundation grant GNSF/ST06/4-098. RE
acknowledges M. K\'eray for patient encouragement and is also
grateful to NSF, Hungary (OTKA, Ref. No. K67746) and the Science and
Technology Facilities Council (STFC), UK for the financial support
received.
\end{acknowledgements}

\bibliographystyle{aps-nameyear}      
\bibliography{example}   
\nocite{*}


\end{document}